\documentclass[aps,prl,twocolumn,superscriptaddress]{revtex4}%
\usepackage{amsfonts}
\usepackage{amsmath}
\usepackage{amssymb}
\usepackage{graphicx}
\usepackage{xcolor}
\usepackage{epstopdf}
\usepackage{bm}%
\begin{document}
\title{Ferromagnetic Impurity Induced Majorana Zero Mode in Iron-Based Superconductor}
\author{Rui Song}
\affiliation{HEDPS, Center for Applied Physics and Technology and School of Physics, Peking
University, Beijing 100871, China}
\affiliation{HEDPS, Center for Applied Physics and Technology and School of Engineering,
Peking University, Beijing 100871, China}
\affiliation{Anhui Key Laboratory of Condensed Matter Physics at Extreme Conditions, High
Magnetic Field Laboratory, HFIPS, Anhui, Chinese Academy of Sciences, and
University of Science and Technology of China, Hefei, China}
\author{Ping Zhang}
\email{zhang_ping@iapcm.ac.cn}
\affiliation{School of Physics and Physical Engineering, Qufu Normal University, Qufu
273165, China}
\affiliation{HEDPS, Center for Applied Physics and Technology and School of Engineering,
Peking University, Beijing 100871, China}
\affiliation{Institute of Applied Physics and Computational Mathematics, Beijing 100088, China}
\author{Xian-Tu He}
\affiliation{HEDPS, Center for Applied Physics and Technology and School of Engineering,
Peking University, Beijing 100871, China}
\affiliation{Institute of Applied Physics and Computational Mathematics, Beijing 100088, China}
\author{Ning Hao}
\email{haon@hmfl.ac.cn}
\affiliation{Anhui Key Laboratory of Condensed Matter Physics at Extreme Conditions, High
Magnetic Field Laboratory, HFIPS, Anhui, Chinese Academy of Sciences, and
University of Science and Technology of China, Hefei, China}

\begin{abstract}
Recent experiments reported the puzzling zero energy modes associated with
ferromagnetic impurities in some iron-based superconductors with topological
band structures.
Here, we show that the sufficiently strong exchange coupling
between a ferromagnetic impurity and substrate can trigger a quantum phase
transition, beyond which, the phase of the topological surface superconducting
order parameter around the impurity acquires a sign-change. In such a case, we
prove that a Kramers degenerate pair of Majorana modes can be induced at the
boundary separating the two sign-change regimes and trapped around the
impurity in the topological surface superconducting state. Furthermore, we
show that our theory can explain the controversial observations and confusing
features of the zero energy modes from recent experiments in some iron-based superconductors.

\end{abstract}
\maketitle

In superconductor, the impurity can induce various quasi-particle states, such
as Yu-Shiba-Rusinov (YSR) state from the classical impurity scattering
potential\cite{YSR-1,YSR-2,YSR-3} and Kondo resonance state from the impurity
in quantum limit\cite{Kando-1}. Through elucidating their properties, one can
obtain much critical information on electron pairing\cite{RMP-1}. Meanwhile,
the quasi-particle state itself can manifest some unexpected behaviors. In
particular, a interstitial iron impurity (IFI) induced robust zero-energy mode
(ZM) has been reported by the scanning tunnelling micropy/spectroscopy (STM/S)
in iron-based superconductor Fe(Te,Se)\cite{MZM-1}. The subsequent studies
have been extended to other iron-based superconductors such as monolayer
Fe(Te,Se)/SrTiO$_{3}$ and LiFeAs, and the similar ZM is also
observed\cite{MZM-2,MZM-3}. Experimentally, the ZM can only be observed on
partial iron impurities and is robust against the external magnetic field, and
the critical temperature is much below the superconducting transition
temperature $T_{c}$. Besides, these materials share a remarkable feature of
possessing topological bands, which implies the ZMs could be Majorana modes.
On the contrary, some revisited studies on Fe(Te,Se) claim the observed ZM is
just trivial YSR states with near-zero-energy electron-like and hole-like
components\cite{NMZM-1,NMZM-2}. Thus, the understandings of the properties and
the mechanism of the ZMs in these iron-based superconductors are still in
debate\cite{MZM-1, MZM-2, MZM-3, NMZM-1, NMZM-2}.

In this work, we first perform the first-principles calculations to
investigate the interaction between IFI and substrate FeSe$_{0.45}$Te$_{0.55}%
$. The numerical results indicate the exchange coupling $J(\mathbf{r},z)$
between magnetic moment of IFI and spin of the 3$d$ electron of FeSe$_{0.45}%
$Te$_{0.55}$ has the form of Friedel-like oscillation with the characteristic
length $a_{0}$ of lattice constant of iron square lattice. The amplitude of
$J(\mathbf{r},z)$ and the magnetic moment of IFI strongly depend on the height
$z$ between IFI and substrate. We further consider the impact of IFI on
topological surface superconducting order parameter $\Delta(\mathbf{r})$ by
solving the Bogoliubov--de Gennes (BdG) equations defined on iron square
lattice with self-consistency. We find that there exists a quantum phase
transition (QPT) at a critical height $z_{c}$, \textit{i.e}. a critical
$J_{c}(\mathbf{r},z_{c})$, beyond which, $\Delta(\mathbf{r})$ change sign in
the $r<a_{0}$ regime. Then, we prove that a Kramers degenerate pair of
Majorana ZMs can be induced at the boundary separating the two sign-change
regimes and trapped by the IFI. For the smaller $J(\mathbf{r},z)<J_{c}%
(\mathbf{r},z_{c})$, the QPT cannot be triggered. The IFI can only induce the
trivial YSR states, which has the near-zero energy in vicinity of QPT. Within
this picture, the contradictions between the results from different STM/S
measurements can be solved, and properties of the ZMs, such as robustness
against external magnetic field and lower critical temperature can also be understood.

\begin{figure}[pt]
\begin{center}
\includegraphics[width=1.0\columnwidth]{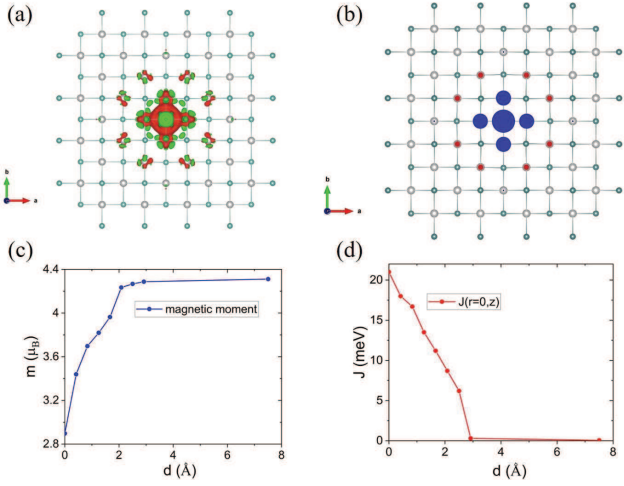}
\end{center}
\caption{(a) The spacial distributions of differential charge density of a
supercell involving a 9$\times$9$\times$1 Fe(Te,Se) substrate and a suspended
IFI. (b) The spacial distributions of the induced spin polarizations of the 3d
electrons of iron atoms in the substrate. The size and color of the dots
denote the strength and direction of the spin polarization, respectively.
(c)-(d) The magnetic moment of IFI and the effective exchange coupling
$J(r=0,z)$ as a function of the height $d$ between the IFI and the substrate,
respectively. }%
\label{figlda}%
\end{figure}

The STM experiment shows that the height of IFI can be tuned by STM
tip\cite{MZM-3,NMZM-2}. During the process of approaching, transition from YSR
states to ZMs happens\cite{MZM-3}. It indicates that the coupling between IFI
and substrate play a crucial role to observe ZMs. To elucidate properties of
such coupling, we construct a 9$\times$9$\times$1 supercell including
substrate Fe(Te,Se) with a suspended IFI. Here, we only summarize main results
in Fig. \ref{figlda}, with calculation details in Ref. \cite{SM}. From Fig.
\ref{figlda} (a), there exists a strong charge transfer between IFI and
substrate iron atoms, and such transfer decays abruptly as expected. The
calculated spin polarizations of substrate shown in Fig. \ref{figlda} (b)
indicate $J(\mathbf{r},z_{0})$ has the form of Friedel-like oscillation, which
is consistent with the neutron scattering experiment on the Fe(Te,Se) with
higher concentration of Te\cite{Neutron}. The characteristic length measured
oscillation period is about lattice constant $a_{0}$ from Fig. \ref{figlda}
(b). The findings are further supported by magnetic moment of IFI as a
function of height, as shown in Fig. \ref{figlda} (c). As IFI approaches the
substrate, the magnetic moment of IFI is suppressed. It indicates spin
transfer also happens and exchange coupling between magnetic moment of IFI and
substrate is strong. The strength of $J(\mathbf{r}=0,z_{0})$ can be roughly
estimated and is shown in Fig. \ref{figlda} (d)\cite{SM}. 

\begin{figure}[pt]
\begin{center}
\includegraphics[width=1.0\columnwidth]{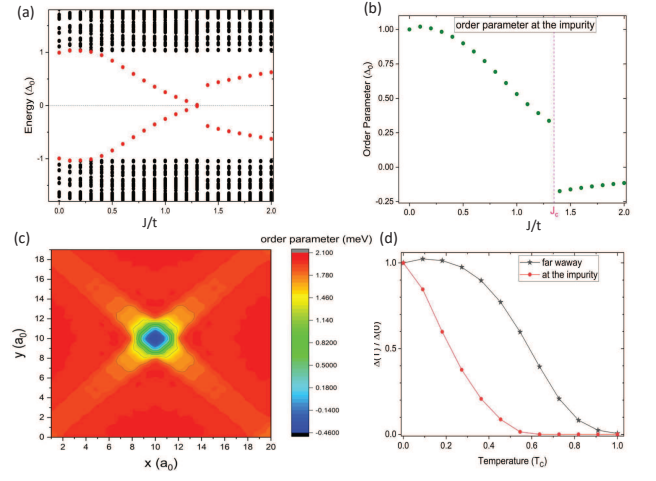}
\end{center}
\caption{(a) The energy spectrum as a function of $J(r=0,z)$ from the
self-consistent solution of Eq. (\ref{Heff}). The red dots with opposite
energy are a pair of YSR states. (b) The order parameter of superconducting
state $\Delta(r=0)$ as a function of $J(r=0,z)$ from the self-consistent
solution of Eq. (\ref{Heff}). (c) The spacial distributions of $\Delta(r)$
under the condition $J(r=0,z)=1.5t$. (d) The temperature evolution of the
$\Delta(r)$ from the self-consistent solution of Eq. (\ref{Heff}). }%
\label{figsfs}%
\end{figure}

Another crucial experimental signature is the presence of a level crossing at
the transition from YSR states to ZMs, and the level crossing is robust
against magnetic field\cite{MZM-3}. This signature indicates the suitable
model related to STS experiments is topological surface Dirac bands with
trivial s-wave pairing\cite{SM}. Thus, we start with such a model defined on
square lattices to evaluate impact of IFI to $\Delta(\mathbf{r})$ of
topological Dirac states on the surface of Fe(Te,Se) substrate. The model
Hamiltonian is,\begin{equation}
H_{eff}=H_{BdG}+H_{coup},\label{Heff}\end{equation}
and\begin{align}
H_{BdG} &  =-\mu\sum_{i}c_{i}^{\dag}c_{i}-it\sum_{<i,j>}c_{i\sigma}^{\dag
}(\mathbf{S}^{\sigma\sigma^{\prime}}\times\mathbf{\hat{d}}_{ij})\mathbf{\mathbf{\cdot\hat{z}}}c_{j\sigma^{\prime}}\nonumber\\
&  +\sum_{i}\Delta_{i}(c_{i\uparrow}^{\dag}c_{i\downarrow}^{\dag
}+c.c.),\label{Hbdg}\\
H_{coup} &  =\int d\mathbf{r}J(\mathbf{r},z_{0})\mathbf{S}_{imp}\mathbf{\cdot\mathbf{\bm{\sigma}}}.\label{Hcoup}\end{align}
Here, $\mu$ is the chemical potential. The second term in Eq. (\ref{Hbdg})
describes the topological surface Dirac states defined on square lattices.
$\mathbf{\hat{d}}_{ij}$ is the unit vector pointing from $i$ to $j$.
$\Delta_{i}$ is site-dependent superconducting order parameter. Note that such
trivial s-wave pairing is good approximation for the topological surface Dirac
state, can give the consistent results with the STS experiments\cite{MZM-3,SM}, and is widely adopted to study the topological properties of the ion-based
superconductors\cite{TS-1, TS-2, TS-3, TS-4, TS-5, Po-2,AV-1, Po-6, Po-7,
Po-8}. $\mathbf{S}_{imp}$ and $\mathbf{\mathbf{\bm{\sigma}}}$ in Eq.
(\ref{Hcoup}) label magnetic moment of IFI and spin of Fe of substrate,
respectively. Here, we only consider $z$-directional spin polarization.
$J(\mathbf{r},z_{0})$ is important only in the first oscillating period
$a_{0}$. The nearest-neighbor $J(\mathbf{r},z_{0})$ is less than $J_{c}$
according to the calculation. Here, we only consider the on-site term
$J(\mathbf{r}=0,z_{0})$ for simplicity\cite{SM}. As $J(\mathbf{r}=0,z_{0})$
increases from zero, there exists a QPT\cite{RMP-1, QPT-1, QPT-2} at a
critical $J_{c}\sim1.3t$, beyond which, $\Delta(r)$ suddenly changes sign and
becomes negative, as shown in Fig. \ref{figsfs} (b). Meanwhile, the level
crossing of two components of YSR states happens, as shown in Fig.
\ref{figsfs} (a). Then, the spacial distributions of $\Delta(r)$ in Fig.
\ref{figsfs} (c) indicate $\Delta(r)$ acquires a $\pi$ phase difference in
$r<R_{0}$ regime in comparison with that in $r>R_{0}$ regime. Note that
$R_{0}$ can take the value of lattice constant $a_{0}$ if the nearest-neighbor
term of $J(\mathbf{r},z_{0})$ is involved\cite{SM}. This is a very crucial
result from the effect of IFI\cite{QPT-1, QPT-2,Order-1,Order-2,Order-3}.
Though QPT is not driven by temperature, increase of temperature could quench
it. Thereafter, we calculates the critical temperature $T_{q}$ of QPT and find
$T_{q}$ is quite lower than bulk superconducting transition temperature
$T_{c}$. We will return to this temperature effect below. 

\begin{figure}[pt]
\begin{center}
\includegraphics[width=1.0\columnwidth]{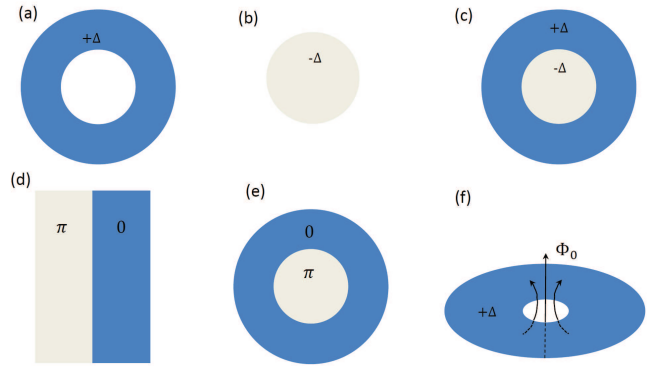}
\end{center}
\caption{(a) The infinite ring with uniform positive superconducting pairing
$\Delta$. (b)The finite disk with uniform negative superconducting pairing
$-\Delta$ (c) Our $0-\pi$ disk junction is the spacial combo of (a) and (b).
(d)-(e) The schematic plotting to show the $0-\pi$ line junction bends to form
a $0-\pi$ disk junction as same as (c). (f) The infinite disk with uniform
positive superconducting pairing $\Delta$, which has a centric hole with a
superconducting flux quanta.}%
\label{figp}%
\end{figure}

Now, we consider the effect of spacial variation of $\Delta(r)$ to topological
surface states. The effective Hamiltonian describing the topological surface
superconductivity with sign-change boundary condition is,%

\begin{equation}
H_{s}=[v_{F}(\mathbf{k}\times\mathbf{\mathbf{\bm{\sigma})\cdot\hat{z}}}%
-\mu]\tau_{z}+\Delta(r)\tau_{x}. \label{Hs}%
\end{equation}
Here, $H_{s}$ is spanned in Nambu space, \textit{i.e.}, $[c_{\uparrow
},c_{\downarrow},c_{\downarrow}^{\dag},-c_{\uparrow}^{\dag}]$. $\tau_{x/z}$ is
Pauli matrix to span particle-hole space. $\Delta(r)=$ $-\Delta_{1}$ when
$r<R_{0}$ and $\Delta(r)=$ $\Delta_{2}$ when $r>R_{0}$ with $\Delta_{1/2}>0$
and $\Delta_{1}<\Delta_{2}$. The phase of $\Delta_{1/2}$ is uniform and is
omitted due to the absence of topological defect such as vortex. Thus,
$\Delta_{1/2}$ is angle-independent and $\Delta(r)$ is real in Eq. (\ref{Hs}).
In continuum limit, eigen-equation of $H_{s}$ is%

\begin{equation}
H_{s}(\mathbf{k}\rightarrow-i\mathbf{\nabla})\psi(r,\theta)=E\psi(r,\theta),
\label{Hs1}%
\end{equation}
which can be solved under boundary conditions with a $0-\pi$ disk junction
shown in Fig. \ref{figp} (c).

Before solving equation (\ref{Hs1}), we give a simple physical picture to
understand existence of a Kramers degenerate pair of Majorana ZMs of the
model. The Hamiltonian in Eq. (\ref{Hs}) preserves particle-hole symmetry
(PHS) with $\mathcal{C}=i\sigma_{y}\tau_{y}K$ and time-reversal symmetry (TRS)
with $\mathcal{T}=i\sigma_{y}\tau_{0}K$. The $0-\pi$ disk junction in Fig.
\ref{figp} (c) can come from the combo of geometries in Figs. \ref{figp} (a)
and (b). We know that both geometries in Figs. \ref{figp} (a) and (b) host
none edge bound states due to Dirac cone itself being two-dimensional boundary
states\cite{SM}. However, when two geometries in Figs. \ref{figp} (a) and (b)
are combined to form $0-\pi$ disk junction in Fig. \ref{figp} (c), edge bound
states must emerge. This behavior can be understood from Fig. \ref{figp} (d)
to (e). $0-\pi$ disk junction in Fig. \ref{figp} (e) can also obtained by
bending $0-\pi$ line junction in Fig. \ref{figp} (d) to connect two ends. It
is well known that $0-\pi$ line junction can support the one-dimensional
linear-dispersion bound states\cite{MZM1D-1, MZM1D-2}. Likewise, $0-\pi$ disk
junction in Figs. \ref{figp} (c) and (e) should also have edge bound states.
Such difference between Figs. \ref{figp} (a) and (b) and Figs. \ref{figp} (c)
and (e) lies in that the wave function $\psi(r,\theta)$ in $0-\pi$ disk
junction in Figs. \ref{figp} (c) and (e) must obey crucial
\textit{antiperiodic boundary condition}, \textit{i.e.}, $\psi(r,\theta)=$
$-\psi(r,\theta+2\pi)$ to get bound states, which is explicitly pointed out by
Fu \textit{et al}\cite{MZM1D-1}. Further considering geometry changes from
Fig. \ref{figp} (d) to (e), the one-dimensional linear-dispersion bound states
have to split into series quantized modes labeled by quantum numbers of
angular momentum, among which, a pair of Majorana ZMs, must emerge. Such
emergence can be understood from vortex case shown in Fig. \ref{figp} (f). The
single-value condition requires wave function in vortex case is
\textit{periodic, i.e., }$\psi_{v}(r,\theta)=\psi_{v}(r,\theta+2\pi)$. If one
does a gauge transformation $\psi_{v}(r,\theta)\rightarrow e^{i\sigma
_{z}\theta/2}\psi_{v}^{\prime}(r,\theta)$, the phase winding of
superconducting pair is eliminated, and new wave function $\psi_{v}^{\prime
}(r,\theta)$ must obey \textit{antiperiodic boundary condition }$\psi
_{v}^{\prime}(r,\theta)=-\psi_{v}^{\prime}(r,\theta+2\pi)$\cite{SM}. It means
that applying magnetic flux is equivalent to changing boundary conditions of
wave function\cite{Anti}. In this sense, our case is equivalent to vortex case
by further taking into account another TR counterpart\cite{SM}. Therefore,
Majorana ZMs must emerge for Hamiltonian in Eq. (\ref{Hs}) with $0-\pi$ disk
junction in Figs. \ref{figp} (c) and (e)\cite{Anti-1}. 

\begin{figure}[pt]
\begin{center}
\includegraphics[width=1.0\columnwidth]{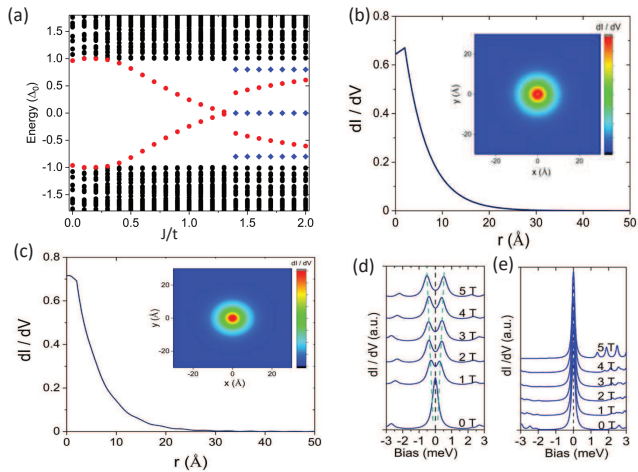}
\end{center}
\caption{(a) The overall energy spectrum from both YSR states (red color) of
IFI scattering and the Majoran ZMs (blue color at zero energy) at the boundary
separated the sing-change regimes as a function of $J(r=0,z)$. The former is
the same as Fig. \ref{figsfs} (a), and the latter is from the numerical
solution of Eq. (\ref{Hs}) defined in the finite disk configuration. (b) The
simulated $dI/dV$ profile for the Majorana zero mode from the analytic
solutions of Eq. (\ref{Hs}) defined in the boundless disk configuration. (c)
The simulated $dI/dV$ profile for the Majorana ZMs by take into account the
quasi-particle scattering. The inset in (b) and (c) are the simulated $dI/dV$
two-dimensional spacial profiles. (d) and (e) The external magnetic field
effect to the YSR states and the Majorna ZMs, respectively. }%
\label{figbrd}%
\end{figure}

The above arguments can be exactly proven by both analytic and numerical
solutions of Eq. (\ref{Hs1})\cite{SM}. For boundless $0-\pi$ disk junction in
Fig. \ref{figp} (c), wave function of the first Majorana ZM takes the form
$\psi_{1}(r,\theta)=[e^{-i\theta/2}u_{\uparrow}(r),e^{i\theta/2}u_{\downarrow
}(r),e^{-i\theta/2}v_{\downarrow}(r),-e^{i\theta/2}v_{\uparrow}(r)]$ with
condition $u_{\sigma}(r)=-$ $v_{\sigma}(r)$. $u_{\sigma}(r)=a_{\sigma}%
J_{\mp1/2}(k_{F}r)e^{r/\xi_{1}}$ for $r<R_{0}$ and $u_{\sigma}(r)=b_{\sigma
}J_{\mp1/2}(k_{F}r)e^{-r/\xi_{2}}$ for $r>R_{0}$. $J_{\mp1/2}(k_{F}r)$ is
Bessel functions with $\mp1/2$ for spin up and down, respectively. $a_{\sigma
}$ and $b_{\sigma}$ are coefficients determined by continuity and
normalization of wave function. Fermi wave vector $k_{F}=\mu/v_{F}$. Decay
length $\xi_{1/2}=v_{F}/\Delta_{1/2}$. The wave function of the second
Majorana ZM can be obtained by $\psi_{2}(r,\theta)=\mathcal{T}$ $\psi
_{1}(r,\theta)$. Note that the mini-gap to protect Majorana ZMs is
proportional to $v_{F}/R_{0}$, which ensures only a pair of Majorana ZMs
survive for small $R_{0}$\cite{SM}. STS measured differential conductance
$dI/dV\propto$ $\sum_{\sigma}r[|u_{\sigma}(r)|^{2}\delta(\omega-eV)+$
$|v_{\sigma}(r)|^{2}\delta(\omega+eV)]$. The case for Majorana ZMs is plotted
in Fig.\ref{figbrd} (b), from which, spacial profile of $dI/dV$ is consistent
with observations in monolayer Fe(Te,Se)/SrTiO$_{3}$\cite{MZM-2} but has
subtle difference near $r=0$ in comparison with the measurements in bulk
Fe(Te,Se) and LiFeAs\cite{MZM-1, MZM-3}. We argue this tiny difference is from
effect of IFI, such as mixture of electronic state, electron's inelastic
tunneling process or finite quasi-particle scattering etc. In Fig.
\ref{figbrd} (c), we consider modulation from finite quasi-particle scattering
and resulting spectrum is quite similar to cases in bulk Fe(Te,Se) and
LiFeAs\cite{SM}. For finite $0-\pi$ disk junction, numerical results are also
consistent with analytic solutions\cite{SM}. 

Aforementioned theory can be utilized to understand multiple features and
common properties of ZMs in iron-based superconductors. We summarize
bound-state spectrum as function as $J(\mathbf{r}=0,z_{0})$ in Fig.
\ref{figbrd} (a). There exists a pair of near-zero-energy YSR states from IFI
when $J(\mathbf{r}=0,z_{0})$ is close to $J_{c}$. The electron-like and
hole-like components of a pair of near-zero-energy YSR states has opposite
spin polarizations. Thus, the whole of them shows no spin-resolved feature.
When $J(\mathbf{r}=0,z_{0})$ is larger than $J_{c}$, a pair of
near-zero-energy YSR states steeply split, and the robust Majorana ZMs emerge
and located at the boundary separated by two sign-change regimes. Note that
the boundary is very close to the IFI. Thus, the contradictions from different
STM/S experiments root in the selected IFIs with different exchange couplings
$J(\mathbf{r}=0,z)$, which coincides with the fact that ZMs can be only
observed on a partial IFIs\cite{MZM-1, MZM-2, MZM-3}. The fragileness of
near-zero-energy YSR state and the robustness of Majorana ZMs against external
magnetic field can also be understood. Consider Zeeman energy $m_{z}$ of
external magnetic field, for a pair of near-zero-energy YSR states, opposite
spin polarization indicates they have to split according to $m_{z}\sigma_{z}$,
as shown in Fig. \ref{figbrd} (d). For Majorana ZMs, the Hamiltonian
$H_{s}(\mathbf{k}\rightarrow-i\mathbf{\nabla})$ defined in $0-\pi$ disk
junction possesses a hidden mirror symmetry $\mathcal{M}_{l=0}=i\sigma_{y}%
\tau_{y}\hat{O}(r)$. $\hat{O}(r)$ is a spacial inverse operator along radial
direction with inverse center at $R_{0}$\cite{SM}. The degeneracy of a pair of
Majorana ZMs is protected by this hidden mirror symmetry against $\sigma_{z}$
and $\sigma_{x}$ Zeeman field. However, in-plane Zeeman term $m_{y}\sigma_{y}$
can split the degeneracy of Majorana ZMs\cite{Po}. This behavior can be
testified by future experimental measurement. Turn to band structure, Zeeman
term $m_{z}$ can be added into $H_{s}$ in Eq. (\ref{Hs}) to open a gap to
Dirac bands. Then, the solution forms of Majorana ZMs are not changed but with
a modulated $k_{F}^{\prime}=\sqrt{\mu^{2}-m_{z}^{2}}/v_{F}$\cite{SM}. Thus,
the Majorana ZMs are robust under condition $m_{z}<$ $\mu$, as shown in Fig.
\ref{figbrd} (e). Note that $H_{s}$ plus $m_{z}\sigma_{z}$ is also one copy of
decoupled Hamiltonian to describe the case in monolayer Fe(Te,Se)/SrTiO$_{3}%
$\cite{MZM-2}, in which, Dirac bands are from the bulk. Some experiments have
observed that Majorana ZMs disappear at a temperature below $T_{c}%
$\cite{MZM-1, MZM-2, MZM-3}. This behavior can also be understood from the
self-consistent calculation results in Fig. \ref{figsfs} (d). The sign-change
$\Delta(r=0)$ decays to zero at about $T_{q}\sim0.6T_{c}$. We argue this is
the primary reason for temperature effect in spite of the possible
quasi-particle poisoning\cite{Po-1, Po-2}. It is worth noting that magnetic
impurity induced robust energy mode has also been observed in PbTaSe$_{2}%
$\cite{PbTaSe}, which also has topological bands and is superconducting.
Within our theory, the observations in PbTaSe$_{2}$ can be well understood.

At last, the reliablity of theory can be enhanced by estimating some relevant
parameters.
The first one is $J_{c}\sim1.3t$ with $t$ measuring the energy
scale of the surface Dirac state, \textit{i.e.}, $t\sim v_{F}k_{F}$. Acoording
to the experiment\cite{Po-2,Pa-1}, $v_{F}\sim$ 250meV\AA , and $k_{F}\sim
$0.02\AA $^{-1}$. Then, $J_{c}\sim$7meV. It is also the reason why single IFI
with quite small exchange coupling can induce the QPT and relevant Majorana
modes. IFI can have a magnetic moment $m\sim$ $5\mu_{B}$, which induces a
magnetic dipole field $B(r)=\mu_{0}m/4\pi r^{3}$\cite{MF-1,MF-2}. The induced
magnetic flux can be calculated by setting lower limit of integral cut-off to
be a Wigner-Seitz radius of square lattice. If one quantized magnetic vortex
emerges, it require the magnetic moment extend to be 10$^{4}\mu_{B}$, which is
only possible for a magnetic cluster in nanoscale.

In conclusion, we provide a new understanding to resolve the debate about
whether STM/S observed ZMs induced by IFI on some iron-based superconductors
is Majorana ZMs or not. We find a QPT can be triggered by the exchange
coupling between the IFI and substrate. Then, the local superconducting order
parameter of the surface superconduting state changes sign around the
impurity, and we prove that a robust Kramers degenerate pair of Majorana ZMs
can be induced and trapped around the IFI. Our theory can explain series
confusing features observed by experiments. More meaningfully, our theory can
be extended to other material categories, which host both topological bands
and superconductivity.

\begin{acknowledgments}
The authors thank J. P. Hu, Z. Fang, C. Fang, X. X. Wu, S. B. Zhang, S. S.
Qin, F. W. Zheng, H. F. Du, L. Shan, Z. Y. Wang, S. C. Yan and X. Y. Hou for
helpful discussions. This work was financially supported by the National Key
R\&D Program of China No. 2017YFA0303201, National Natural Science Foundation
of China under Grants (No. 12022413, No. 11674331, No.11625415), the
\textquotedblleft Strategic Priority Research Program (B)\textquotedblright%
\ of the Chinese Academy of Sciences, Grant No. XDB33030100, the `100 Talents
Project' of the Chinese Academy of Sciences, the Collaborative Innovation
Program of Hefei Science Center, CAS (Grants No. 2020HSC-CIP002), the CASHIPS
Director's Fund (BJPY2019B03), the Science Challenge Project under Grant No.
TZ2016001, the Major Basic Program of Natural Science Foundation of Shandong
Province (Grant No. ZR2021ZD01). A portion of this work was supported by the
High Magnetic Field Laboratory of Anhui Province, China.
\end{acknowledgments}

\begin{widetext}
\section{DFT calculating methods}

First-principles calculations were performed by density functional theory
(DFT) using the Vienna ab initio simulation package (VASP) \cite{vasp1,vasp2}.
The plane-wave basis with an energy cutoff of 350 eV was adopted. The
electron-ion interactions were modeled by the projector augmented wave
potential (PAW) \cite{PAW} and the exchange-correlation functional was
approximated by the Perdew-Burke-Ernzerhof-type (PBE) generalized gradient
approximation (GGA) \cite{PBE}. Here we use the parameter "I-CONSTRAINED-M" in
VASP to constrain the direction of the Fe impurity always along the z axis and
give the substrate a very weak magnetic background. As shown in Fig.
\ref{charge}, the differential charge density can reflect the coupling between
the impurity and substrate, as the height increases, the chemical adsorption
will turn into physical adsorption. To estimate the value of $J(r,z)$, we can
set the impurity having a FM and AFM coupling with its nearest iron atoms
respectively. And using a simple Heisenberg Hamiltonian, we can calculate the
strength of exchange coupling. The local Hamiltonian reads \begin{figure}[h]
\centering
\par
\includegraphics[width=120mm]{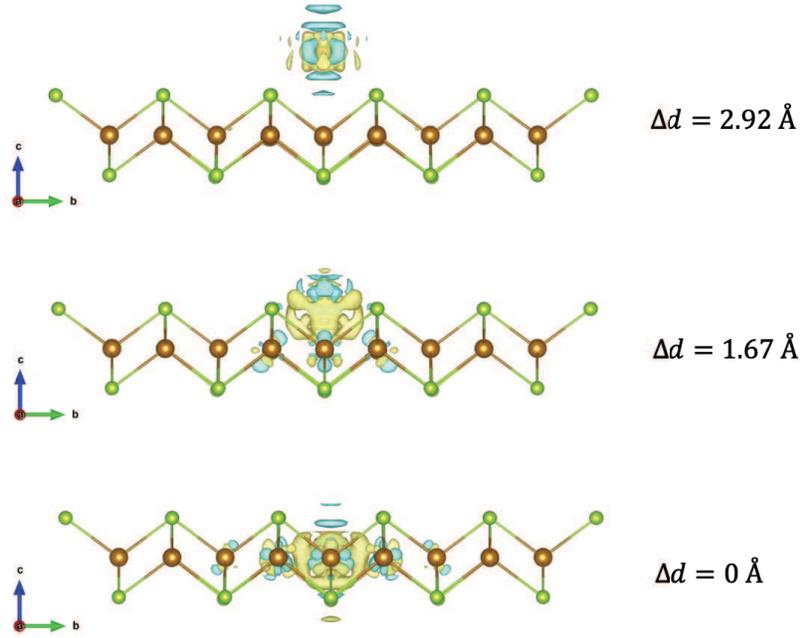}\newline\caption{Differential charge
density of different height (a) 2.92\AA , (b) 1.67\AA \ and (c) 0\AA .}%
\label{charge}%
\end{figure}%

\begin{equation}
H_{M}=-J_{H} \sum_{NN}S_{imp}\cdot S_{NN}, \label{Heisenberge}%
\end{equation}
where $S_{imp}$ and $S_{NN}$ is the magnetic moment of the impurity and its
nearest neighbor iron atoms. Because the impurity has four nearest neighbor,
we can get
\begin{equation}
\Delta E = H_{AFM}-H_{FM} =8 J_{H} S_{imp} S_{NN}. \label{energy}%
\end{equation}
The effective magnetic field generated by the impurity is
\begin{equation}
h_{imp}=\frac{n\langle S_{imp}\rangle}{\mu_{B}}\int J_{H}(r)d \mathbf{r}
\label{eff-B}%
\end{equation}
where $n$, $S_{imp}$,$\mu_{B}$ labels the concentration of localized moments,
the average value of the localized spins, Bohr magneton and ferromagnetic
exchange integral, respectively \cite{RMP}. For a single impurity, we have
$n=1$ and $J_{H}(r)=J_{H}\delta(r)$. The coupling Hamiltonian is
\begin{equation}
H_{coup}=\int d\mathbf{r}J(\mathbf{r},z_{0})\mathbf{S}_{imp}\cdot
\mathbf{\sigma}%
\end{equation}
The exchange interaction is very local, so we can assume that $J(\mathbf{r}%
,z_{0})=J \delta(r)$, and then we can adopt a mean-field approximation as
\begin{equation}
H_{coup}=\langle J S_{imp}\rangle\sum\limits_{\sigma}\sigma c_{i_{0}}^{\dag
}c_{i_{0}}. \label{mean-exchange}%
\end{equation}
Comparing Eq. \ref{mean-exchange} and Eq. \ref{eff-B}, we find that the
existence of the magnetic impurity is equivalent to applying a local magnetic
field at the impurity site. So the strength of the exchange coupling can be
characterized by the effective magnetic field generated by the impurity, which
can be calculated as
\begin{equation}
\langle J S_{imp}\rangle=\frac{\Delta E}{8 S_{NN}} \label{J-app}%
\end{equation}
In our calculations, the $S_{NN}$ is set to be $0.05\mu_{B}$, $S_{imp}=\pm
3\mu_{B}$ for FM and AFM coupling respectively.

\section{BdG method in lattice model}

\begin{figure}[h]
\centering
\par
\includegraphics[width=100mm]{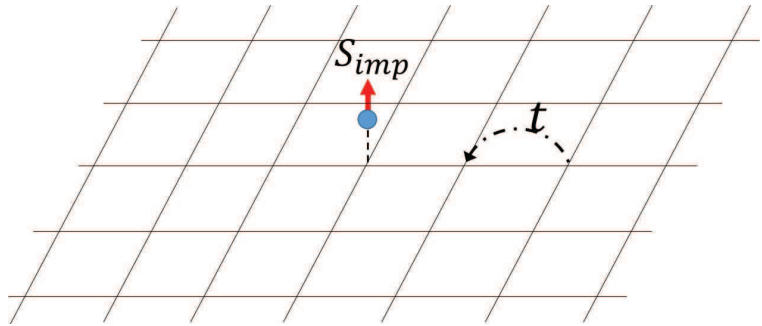} \caption{Scheme of the lattice
model.}%
\label{square}%
\end{figure}To model the superconductivity in right way, it is better to start
from the experimental measurements. From Fig. 3 in the experimental
paper\cite{EXP}, one can find that the two YSR states always cross at a
critical value of coupling between the impurity and substrate. Note that
applying magnetic field is equivalent to change the coupling between the
impurity and substrate and just move the crossing point. This robust crossing
feature is crucial to decide how to model the superconductivity in right way.
There are two possibilities to model the superconductivity related to the
STM/STS experiments. i.e., the topological surface Dirac bands plus trivial
s-wave pairing and trivial bulk multi-bands plus S$_{\pm}$ pairing. In the
former one, the superconductivity of the surface Dirac bands is from the
self-proximity effect or inter-band scattering around $\bar{\Gamma}$ point in
the surface Brillouin zone. The trivial bulk bands around $\bar{M}$ point have
no effect to the surface Dirac bands near $\bar{\Gamma}$ point. Thus, the
superconductivity of the surface Dirac bands near $\bar{\Gamma}$ point can be
reduced to have the trivial s-wave pairing, if the surface Dirac bands play a
dominated role. We calculate the impurity induced YSR states for both models
to see which one is consistent with the experimental measurements.

\subsection{Topological surface Dirac model with trivial s-wave pairing}

We consider a square lattice, the topological surface Dirac Hamiltonian can be
expressed as
\begin{equation}
H_{surf}=-\sum_{i}\mu c_{i}^{\dagger}c_{i}-it\sum_{\langle i,j\rangle
}c_{i\sigma}^{\dagger}(\mathbf{S}^{\sigma\sigma^{\prime}}\times\mathbf{\hat
{d}}_{ij})\cdot\mathbf{\hat{z}}c_{j\sigma^{\prime}} \label{self-consistent}%
\end{equation}
which includes the on-site term, and linear Dirac term. Now a magnetic
impurity is put onto the site $i_{0}$ and thus would introduce an exchange
term as
\begin{equation}
H_{coup}=-J(c_{i_{0}\uparrow}^{\dagger}c_{i_{0}\uparrow}-c_{i_{0}\downarrow
}^{\dagger}c_{i_{0}\downarrow}), \label{self-consistent1}%
\end{equation}
which performs like an effective magnetic field generated by the impurity.
Note that $S_{imp}$ is merged into $J$ for simplicity. And when
superconductivity is induced, the surface trivial s-wave Cooper pairing
potential is introduced as
\begin{equation}
H_{\Delta}=\sum_{i}(\Delta c_{i\uparrow}^{\dagger}c_{i\downarrow}^{\dagger
}+h.c.). \label{self-consistent2}%
\end{equation}
The total Hamiltonian is
\begin{equation}
H=H_{surf}+H_{coup}+H_{\Delta}. \label{total}%
\end{equation}

We can perform the Bogoliubov transformation
\begin{equation}
c_{i\sigma}=\sum_{n}^{^{\prime}}(u_{i\sigma}^{n}\gamma_{n}-\sigma v_{i\sigma
}^{n\ast}\gamma_{n}^{\dagger})\label{self-consistent3}%
\end{equation}
where $^{\prime}$ denotes summation over the positive eigenvalues, and
numerically solve the equations
\begin{equation}
\sum_{j}\hat{H}_{ij}\phi_{j}=E_{n}\phi_{i}%
\end{equation}
in the Nambu spinor representation $\phi_{i}=(u_{i\uparrow}^{n},u_{i\downarrow
}^{n},v_{i\uparrow}^{n},v_{i\downarrow}^{n})^{T}$ \cite{numerical-2}. The
matrix elements read
\begin{equation}
\hat{H}_{ij}=\left(
\begin{array}
[c]{cccc}%
\Lambda_{ij\uparrow} & T_{ij}^{\uparrow\downarrow} & 0 & \Delta_{i}\\
T_{ij}^{\downarrow\uparrow} & \Lambda_{ij\downarrow} & \Delta_{i} & 0\\
0 & \Delta_{i} & -(\Lambda_{ij\uparrow})^{\ast} & (T_{ij}^{\uparrow\downarrow
})^{\ast}\\
\Delta_{i} & 0 & (T_{ij}^{\downarrow\uparrow})^{\ast} & -(\Lambda
_{ij\downarrow})^{\ast}%
\end{array}
\right)  ,
\end{equation}
where $\Lambda_{ij\sigma}=-(\mu+\sigma J\delta_{ii_{0}})\delta_{ij}$.
$T_{ij}^{\sigma\sigma^{\prime}}=-it(\mathbf{S}^{\sigma\sigma^{\prime}}%
\times\mathbf{\hat{d}}_{ij})\cdot\mathbf{\hat{z}}$. And the order parameter
$\Delta_{i}$ should be self-consistently determined as
\begin{equation}
\Delta_{i}(T)=\frac{g}{2}\sum_{n}^{^{\prime}}(u_{i\uparrow}^{n}v_{i\downarrow
}^{n\ast}+u_{i\downarrow}^{n}v_{i\uparrow}^{n\ast})\tanh(\frac{E_{n}}{k_{B}T})
\end{equation}
To get the numerical results, we have adopted a $25\times25$ lattice. The
hopping parameter $t=10\ meV$, homogeneous order parameter $\Delta_{0}=2\ meV$
and chemical potential $\mu=0.1t$. The calculated spectrum of $H$ in Eq.
\ref{total} is shown in Fig. \ref{op-cal} (a) and (b), from which, one can
find there exist a robust level crossing for two YSR states against the
external magnetic field. The level crossing is schematically shown in Fig.
\ref{op-cal} (d). The existence of level crossing is consistent with the
experimental measurements, and indicates a QPT happens. At the QPT, the order
parameter at the impurity site is discontinuous and suddenly change sign as
shown in Fig. \ref{op-cal} (c).

\begin{figure}[h]
\centering
\par
\includegraphics[width=150mm]{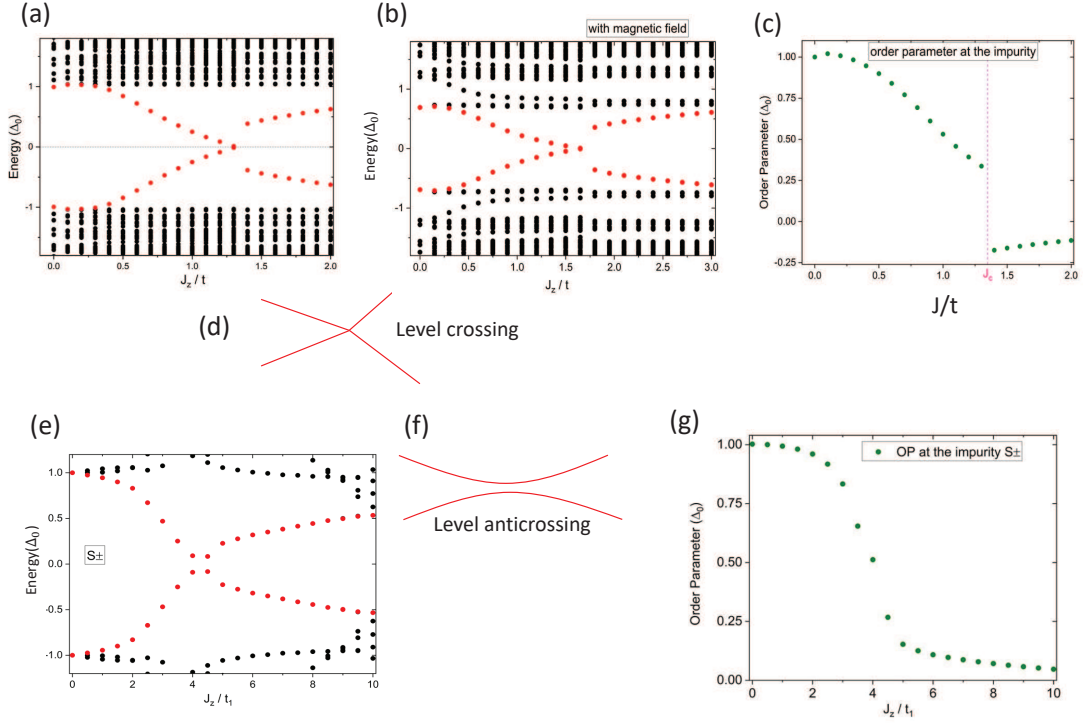} \caption{(a)-(c) The energy
spectrum and superconducting order parameter of topological surface Dirac
superconducting state with magnetic impurity scattering described by
Hamiltonian $H$ in Eq. \ref{total}. (a) without magnetic field and (b) with
magnetic field. In (c), the superconducting order parameter at the impurity
site as a function of $J$. (d) The scheme of level crossing. (e)-(g) The
energy spectrum and superconducting order parameter of two-band model with
S$_{\pm}$ pairing and magnetic impurity scattering described by Hamiltonian
$H^{\prime}$ in Eq. \ref{total-1}. (e) without magnetic field. (f) The scheme
of level anticrossing. (g) the superconducting order parameter at the impurity
site as a function of $J$.}%
\label{op-cal}%
\end{figure}

\subsection{Trivial bulk Multi-bands model with S$_{\pm}$ pairing}

Now we will check the same process between the magnetic impurity and bulk
state of iron-based superconductor. The main difference is the pairing
pattern. The numerical calculation is also carried on a square lattice and the
minimal two-orbital band structure is adopt\cite{twoband}. The two-orbitals
model is described by%

\[
H_{b}=\sum_{k}\psi_{k}^{\dag}[(\epsilon_{+}(k)-\mu)+\epsilon_{-}(k)\tau
_{z}+\epsilon_{xy}(k)\tau_{x}]\psi_{k},
\]
with $\psi_{k}^{\dag}=[d_{xz}^{\dag}(k),d_{yz}^{\dag}(k)],$ and%

\begin{align*}
\epsilon_{\pm}(k)  &  =\frac{\epsilon_{x}(k)\pm\epsilon_{y}(k)}{2}\\
\epsilon_{x}(k)  &  =-2t_{1}\cos k_{x}-2t_{2}\cos k_{y}-4t_{3}\cos k_{x}\cos
k_{y}\\
\epsilon_{y}(k)  &  =-2t_{2}\cos k_{x}-2t_{1}\cos k_{y}-4t_{3}\cos k_{x}\cos
k_{y}\\
\epsilon_{xy}(k)  &  =-4t_{4}\sin k_{x}\sin k_{y}%
\end{align*}
After introducing the interacting term and at the mean field approximation, we
obtain the BdG Hamiltonian as%

\[
H_{b}+H_{\Delta}^{\prime}=-\sum_{ij,\alpha\beta}(t_{ij,\alpha\beta}+\mu
\delta_{ij}\delta_{\alpha\beta})d_{i\alpha}^{\dagger}d_{j\beta}+\sum
_{<<i,j>>}(\Delta_{ij,\alpha\beta}d_{i\alpha\uparrow}^{\dagger}d_{j\beta
\downarrow}^{\dagger}+h.c.).
\]
Here, $\Delta_{ij,\alpha\beta}$ describes the next-nearest neighbor pairing
with S$_{\pm}$ symmetry \textit{i.e.}, $\cos k_{x}\cos k_{y}$ form in momentum
space. The IFI induced intra-orbital scattering is considered as
\[
H_{coup}^{\prime}=J\sum_{\alpha}(d_{i_{0}\alpha\uparrow}^{\dagger}%
d_{i_{0}\alpha\uparrow}-d_{i_{0}\alpha\downarrow}^{\dagger}d_{i_{0}%
\alpha\downarrow}).
\]

The total Hamiltonian is%

\begin{equation}
H^{\prime}=H_{b}+H_{\Delta}^{\prime}+H_{coup}^{\prime} \label{total-1}%
\end{equation}

The self-consistent condition is%

\[
\Delta_{ij}(T)=\frac{g}{2}\sum_{<<i,j>>n}^{^{\prime}}(u_{i\uparrow}%
^{n}v_{j\downarrow}^{n\ast}+u_{i\downarrow}^{n}v_{j\uparrow}^{n\ast}%
)\tanh(\frac{E_{n}}{k_{B}T}),
\]
and the final homogeneous order parameter is expressed as $\bar{\Delta}%
_{i}=\frac{1}{8}\sum_{j,\alpha\beta}\Delta_{ij}^{\alpha/\beta}.$The calculated
spectrum of $H^{\prime}$ in Eq. \ref{total-1} is shown in Fig. \ref{op-cal}
(e), from which, one can find there only exist a level anticrossing for two
YSR states. The level anticrossing is schematically shown in Fig. \ref{op-cal}
(f). This feature is inconsistent with the experimental measurements, and no
QPT happens. The order parameter at the impurity site continuously decays to
zero as $J$ change, as shown in Fig. \ref{op-cal} (g).

By comparing the calculation results of the above two models with the
experimental measurements\cite{EXP}, we arrive the right model related with
the STS experiments is the topological surface Dirac model with trivial s-wave
pairing. We start with this model in the main text.

\section{The equivalence between vortex case and impurity case}

\subsection{Numerical proof for the antiperiodic boundary condition for the
impurity case}

\begin{figure}[h]
\centering
\par
\includegraphics[width=150mm]{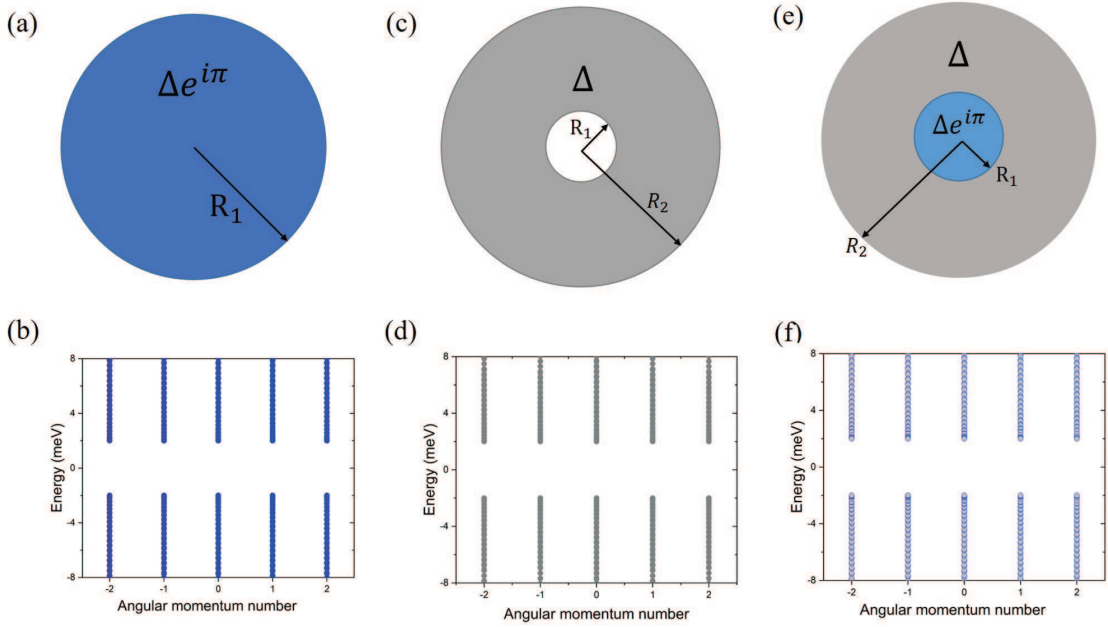} \caption{Schematic diagram of (a) A
$\pi$-phase disk and (c) A 0-phase ring, (e) A combo of (a) and (b). (b), (d)
and (e) The energy spectrum with a periodic boundary condition for (a), (c)
and (e), respectively.}%
\label{periodic}%
\end{figure}

We first calculate the spectrum of the geometries, as shown in Fig.
\ref{periodic}(a), (c) and (e) with periodic boundary condition \textit{i.e}.
$\psi(r,\theta)=\psi(r,\theta+2\pi)$. The numerical results are shown in Fig.
\ref{periodic}(b), (d) and (e), respectively. All $\pi$-phase disk and 0-phase
ring and $0-\pi$ disk junction have no in-gap bound states. The numerical
result for the $0-\pi$ disk junction is not consistent with the well-know
$0-\pi$ line junction results. Therefore, the periodic boundary condition is
not right for the $0-\pi$ disk junction, and the \textit{antiperiodic boundary
condition} \textit{i.e}. $\psi(r,\theta)=-\psi(r,\theta+2\pi)$ has to be used
for the $0-\pi$ junction disk in impurity case.

\subsection{Vortex case}

For comparison, we also analyzed the general vortex-induced Majorana mode.
When there exist quantum vortexes, the SC order parameter should have a
attached phase as $\Delta=\Delta(r) e^{i \theta}$. The complete Hamiltonian
can be expressed as
\begin{equation}
\left(
\begin{array}
[c]{cccc}%
\setlength{\arraycolsep}{0.01pt} -\mu & -e^{-i \theta}v_{F}(\partial_{r}%
-\frac{i}{r}\partial_{\theta}) & \Delta(r)e^{i \theta} & 0\\
e^{i \theta}v_{F}(\partial_{r}+\frac{i}{r}\partial_{\theta}) & -\mu & 0 &
\Delta(r)e^{i \theta}\\
\Delta(r)e^{-i \theta} & 0 & \mu & e^{-i \theta}v_{F}(\partial_{r}-\frac{i}%
{r}\partial_{\theta})\\
0 & \Delta(r)e^{-i \theta} & -e^{i \theta}v_{F}(\partial_{r}+\frac{i}%
{r}\partial_{\theta}) & \mu
\end{array}
\right)  \label{vortex-condition}%
\end{equation}
Correspondingly, the trivial wave functions should have this form
\begin{equation}
\label{vortex-wave}\psi(r,\theta)= e^{i l \theta}\left(
\begin{array}
[c]{c}%
u_{\uparrow}(r)\\
e^{i\theta}u_{\downarrow}(r)\\
e^{-i\theta}v_{\downarrow}(r)\\
v_{\uparrow}(r)
\end{array}
\right)
\end{equation}
Also, the Majorana condition requests that
\begin{align}
l  &  = 0\\
u_{\uparrow}(r)  &  =-\eta v_{\uparrow}(r)\\
u_{\downarrow}(r)  &  =\eta v_{\downarrow}(r)
\end{align}
The radial equation can be obtained as
\begin{equation}
\label{vortex-condition-radial}\left(
\begin{array}
[c]{cccc}%
\setlength{\arraycolsep}{0.01pt} -\mu & -v_{F}(\partial_{r}+\frac{l+1}{r}) &
\Delta(r) & 0\\
v_{F}(\partial_{r}-\frac{l}{r}) & -\mu & 0 & \Delta(r)\\
\Delta(r) & 0 & \mu & v_{F}(\partial_{r}+\frac{l}{r})\\
0 & \Delta(r) & -v_{F}(\partial_{r}-\frac{l-1}{r}) & \mu
\end{array}
\right)  \left(
\begin{array}
[c]{c}%
u_{\uparrow}(r)\\
u_{\downarrow}(r)\\
v_{\downarrow}(r)\\
v_{\uparrow}(r)
\end{array}
\right)  = E \left(
\begin{array}
[c]{c}%
u_{\uparrow}(r)\\
u_{\downarrow}(r)\\
v_{\downarrow}(r)\\
v_{\uparrow}(r)
\end{array}
\right)
\end{equation}
Here, we first derive the analytic solution for Majorana mode when $\Delta(r)$
is approximated as a constant $\Delta_{0}$. Similarly, the equation can be
simplified as
\begin{equation}
u_{\uparrow}^{^{\prime\prime}}(r)+(\frac{2}{\xi_{0}}+\frac{1}{r})u_{\uparrow
}^{^{\prime}}(r)+(\frac{1}{\xi_{0}^{2}}+\frac{1}{r \xi_{0}}+k_{F}%
^{2})u_{\uparrow}(r)=0
\end{equation}
the solution can be obtained as :
\begin{align}
u_{\uparrow}(r)  &  =c_{1}J_{0}(k_{F}r)e^{-r/\xi_{0}}\\
u_{\downarrow}(r)  &  =-c_{1}J_{1}(k_{F}r)e^{-r/\xi_{0}},
\end{align}
which is consistent with Ref. \cite{spin-reso-theo, spin-reso-exp}.
\begin{figure}[h]
\centering
\par
\includegraphics[width=150mm]{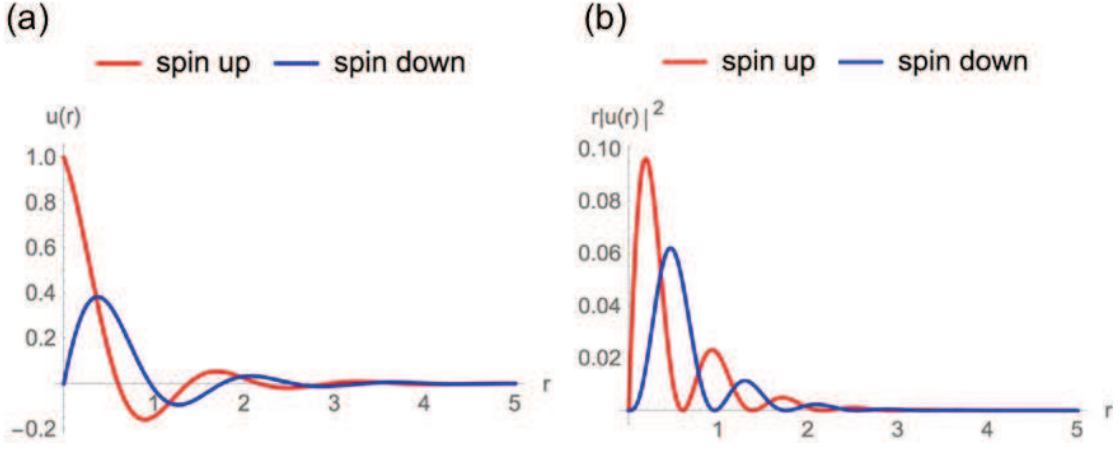}\newline\caption{(a) wave functions
and (b) the probability density for the vortex-induced Majorana mode.}%
\label{wave-function}%
\end{figure}

Now we will perform the numerical solution of Eq.
(\ref{vortex-condition-radial}), the method is the same as above. The space
variation of order parameter is adopted as $\Delta(r)=\Delta_{0}\tanh\frac
{r}{\xi_{0}}$. The results are shown in Fig. \ref{numerical-vortex}.
\begin{figure}[h]
\centering
\par
\includegraphics[width=170mm]{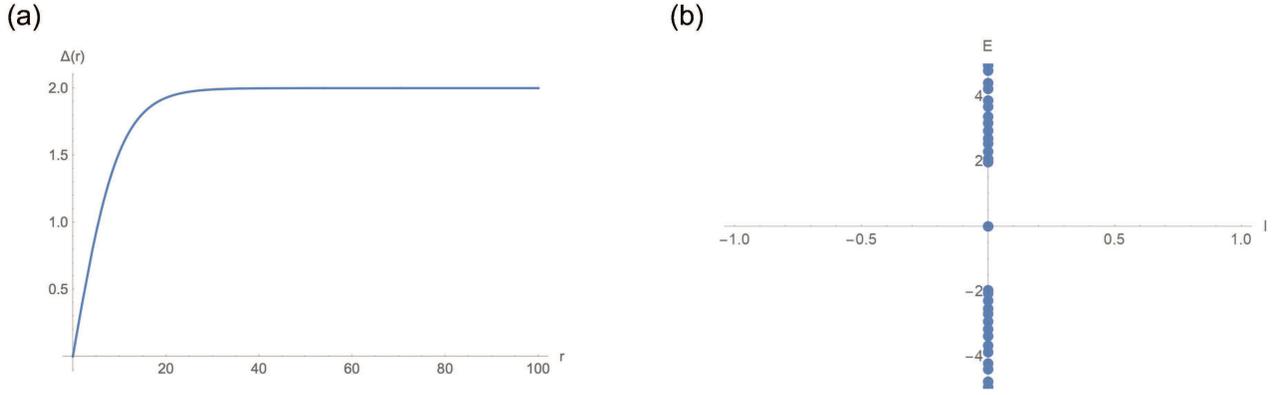}\newline\caption{(a) The order
parameter from the core to the edge of a vortex and (b) the energy spectrum of
momentum $l=0$.}%
\label{numerical-vortex}%
\end{figure}For the vortex condition, its Hamiltonian can be expressed as
\begin{equation}
H_{\nu}=[v_{F}(k_{x}\sigma_{y}-k_{y}\sigma_{x})-\mu]\tau_{z}+\Delta_{0}%
e^{i\nu\theta}\tau_{+}+\Delta_{0}e^{-i\nu\theta}\tau_{-}, \label{vor-H}%
\end{equation}
where $\nu=\pm1$ for (anti-) vortex, $\tau_{\pm}=\tau_{x}\pm i\tau_{y}$. The
wave functions of a vortex has a form shown in Eq. (\ref{vortex-wave}), we
name it $\psi_{v+}$. And the wave function of an anti-vortex can be easily
obtained as
\begin{equation}
\psi_{v-}=e^{il\theta}\left(
\begin{array}
[c]{c}%
e^{-i\theta}u_{\uparrow}(r)\\
u_{\downarrow}(r)\\
v_{\downarrow}(r)\\
e^{i\theta}v_{\uparrow}(r)
\end{array}
\right)  . \label{anti-vortex}%
\end{equation}
Both $\psi_{v+}$ and $\psi_{v-}$ are periodic. However we can apply this gauge
transformation
\begin{equation}
\psi_{v\pm}^{\prime}=U_{\pm}\psi_{v\pm}, \label{gauge-tran}%
\end{equation}
here $U_{+}=diag(e^{-i\theta/2},e^{-i\theta/2},e^{i\theta/2},e^{i\theta/2})$
and $U_{-}=U_{+}^{\ast}$. It is obvious that the new wave functions
$\psi_{v\pm}^{\prime}$ is anti-periodic, the corresponding Hamiltonian can be
obtained as $H_{v\pm}^{\prime}=U_{\pm}H_{\nu}U_{\pm}^{-1}$, which reads
\begin{equation}
H_{v+}^{\prime}=H_{v-}^{\prime}=[v_{F}(k_{x}\sigma_{y}-k_{y}\sigma_{x}%
)-\mu]\tau_{z}+\Delta_{0}\tau_{x}. \label{uniform}%
\end{equation}
Eq. (\ref{uniform}) has the same form with Eq. (\ref{1}) when $B=0$. So its
proved that the bound state in our $0-\pi$ disk junction is equivalent to the
vortex condition, the difference is that in our $0-\pi$ disk junction model
the time-reversal symmetry is preserved so that the Majorana mode is a helical
mode instead of a chiral mode. That can also be reflected by the numerical
calculated results.

\section{Solution of Majorana zero Mode}

\subsection{Analytic solution}

The Hamiltonian of a magnetic field applied to a superconductor with a
Dirac-type topological surface state can be expressed as $H=\int d^{2}%
r\Psi^{\dag}(r)\mathcal{H}\Psi(r)$, where
\begin{equation}
\mathcal{H}=\left[  v_{F}(k_{x}\sigma_{y}-k_{y}\sigma_{x})-\mu\right]
\tau_{z}+\alpha\ \sigma\cdot\mathbf{B}+\Delta(r)\tau_{x}. \label{1}%
\end{equation}
Here $\Psi=(c_{\uparrow},c_{\downarrow},c_{\downarrow}^{\dag},-c_{\uparrow
}^{\dag})$ denotes the Nambu basis, $\sigma_{i}$ and $\tau_{i}$ $(i=x,y,z)$
are Pauli matrices but spans spin and Nambu space respectively, $\Delta(r)$
takes the real value due to no superconducting vortex. $\mathbf{B}%
(\mathbf{r})$ is the external magnetic field $\mathbf{B}=(B_{x},B_{y},B_{z})$.
Due to the existence of impurity, translation invariance is broken. Thus we
need to solve Eq. (1) in real space with cylindrical coordinate system, and we
have following transformation:
\begin{align*}
k_{x}-ik_{y}  &  =-ie^{-i\theta}\partial_{r}-\frac{e^{-i\theta}}{r}%
\partial_{\theta}\\
k_{x}+ik_{y}  &  =-ie^{i\theta}\partial_{r}+\frac{e^{i\theta}}{r}%
\partial_{\theta}%
\end{align*}
Substituting it into Eq. (1), we get the complete form BdG Hamiltonian as%

\begin{equation}
\left(
\begin{array}
[c]{cccc}%
\setlength{\arraycolsep}{0.01pt}\alpha B_{z}-\mu & -e^{-i\theta}v_{F}%
(\partial_{r}-\frac{i}{r}\partial_{\theta}) & \Delta(r) & 0\\
e^{i\theta}v_{F}(\partial_{r}+\frac{i}{r}\partial_{\theta}) & -\alpha
B_{z}-\mu & 0 & \Delta(r)\\
\Delta(r) & 0 & \alpha B_{z}+\mu & e^{-i\theta}v_{F}(\partial_{r}-\frac{i}%
{r}\partial_{\theta})\\
0 & \Delta(r) & -e^{i\theta}v_{F}(\partial_{r}+\frac{i}{r}\partial_{\theta}) &
-\alpha B_{z}+\mu
\end{array}
\right)  . \label{3}%
\end{equation}
Here, we assume the magnetic field is along the z direction. The BdG equation
is
\begin{equation}
\mathcal{H}\psi(r,\theta)=E\psi(r,\theta). \label{4}%
\end{equation}
Assuming the trivial wave function with the \textit{antiperiodic boundary
condition} has the form
\begin{equation}
\psi(r,\theta)=e^{il\theta}\left(
\begin{array}
[c]{c}%
e^{-i\theta/2}u_{\uparrow}(r)\\
e^{i\theta/2}u_{\downarrow}(r)\\
e^{-i\theta/2}v_{\downarrow}(r)\\
-e^{i\theta/2}v_{\uparrow}(r)
\end{array}
\right)  \label{ZMS-1}%
\end{equation}
Then we can obtain the radial equation of Eq. (\ref{4}) as:
\begin{align}
(\alpha B_{z}-\mu)u_{\uparrow}(r)-v_{F}(\partial_{r}+\frac{l+\frac{1}{2}}%
{r})u_{\downarrow}(r)+\Delta(r)v_{\downarrow}(r)  &  =Eu_{\uparrow
}(r)\label{an-1}\\
v_{F}(\partial_{r}-\frac{l-\frac{1}{2}}{r})u_{\uparrow}(r)-\left[  \alpha
B_{z}+\mu\right]  u_{\downarrow}(r)-\Delta(r)v_{\uparrow}(r)  &
=Eu_{\downarrow}(r)\label{an-2}\\
\Delta(r)u_{\uparrow}(r)+(\alpha B_{z}+\mu)v_{\downarrow}(r)-v_{F}%
(\partial_{r}+\frac{l+\frac{1}{2}}{r})v_{\uparrow}(r)  &  =Ev_{\downarrow
}(r)\label{an-3}\\
\Delta(r)u_{\downarrow}(r)-v_{F}(\partial_{r}-\frac{l-\frac{1}{2}}%
{r})v_{\downarrow}(r)+(\alpha B_{z}-\mu)v_{\uparrow}(r)  &  =-Ev_{\uparrow}(r)
\label{an-4}%
\end{align}
The matrix form is :%

\begin{equation}
\left(
\begin{array}
[c]{cccc}%
\setlength{\arraycolsep}{0.01pt}\alpha B_{z}-\mu & -v_{F}(\partial_{r}%
+\frac{\nu+1}{r}) & \Delta(r) & 0\\
v_{F}(\partial_{r}-\frac{\nu}{r}) & -\alpha B_{z}-\mu & 0 & \Delta(r)\\
\Delta(r) & 0 & \alpha B_{z}+\mu & v_{F}(\partial_{r}+\frac{\nu+1}{r})\\
0 & \Delta(r) & -v_{F}(\partial_{r}-\frac{\nu}{r}) & -\alpha B_{z}+\mu
\end{array}
\right)  , \label{radiu}%
\end{equation}
where $\nu=l-\frac{1}{2}$. The Majorana solution requests $\mathcal{C}%
\psi(r,\theta)=\eta\psi(r,\theta)$, where the particle-hole operator
$\mathcal{C}=\tau_{y}\sigma_{y}K$, with $K$ the complex conjugation operator,
and $\eta$ is some constant. Without loss of generality, we assume
$u_{\uparrow(\downarrow)}(r)$ and $v_{\uparrow(\downarrow)}(r)$ is real. This
yields the following constraint condition:
\begin{align}
u_{\uparrow}(r)  &  =\eta v_{\uparrow}(r)\\
u_{\downarrow}(r)  &  =\eta v_{\downarrow}(r)\\
l  &  =0
\end{align}
Obviously, $\eta$ can only be $+1$ or $-1$, and they correspond to the
exponential increase and decay solution, respectively. For an infinite large
2D space, only $\eta=-1$ is reasonable. And then the equation set of Eq.
(\ref{an-1})-(\ref{an-4}) can be reduced as
\begin{align}
(\alpha B_{z}-\mu)u_{\uparrow}(r)-\left[  v_{F}(\partial_{r}+\frac{1}%
{2r})+\Delta(r)\right]  u_{\downarrow}(r)  &  =Eu_{\uparrow}(r),
\label{eeq-1}\\
\left[  v_{F}(\partial_{r}+\frac{1}{2r})+\Delta(r)\right]  u_{\uparrow
}(r)-\left[  \alpha B_{z}+\mu\right]  u_{\downarrow}(r)  &  =Eu_{\downarrow
}(r), \label{eeq-2}%
\end{align}
For the Majorana condition, we can set $E=0$.

Now, let's check the simplest condition where $B_{z}=0$, which means there is
no magnetic effect and then the system should return to the TI+SC model,
however with a antiperiodic boundary condition. When $B_{z}=0$, Eq. (12) and
(13) can be reduced to :
\begin{equation}
u_{\uparrow(\downarrow)}^{\prime\prime}(r)+\frac{u_{\uparrow(\downarrow
)}^{\prime}(r)}{r}+(\frac{\mu^{2}+\Delta_{0}^{2}}{v_{F}^{2}}-\frac{1}{4r^{2}%
})u_{\uparrow(\downarrow)}(r)+\frac{2\Delta_{0}}{v_{F}}\left[  u_{\uparrow
(\downarrow)}^{\prime}(r)+\frac{1}{2r}u_{\uparrow(\downarrow)}(r)\right]  =0
\label{Bessel}%
\end{equation}
The solution of $u_{\uparrow}(r)$ is
\begin{equation}
u_{\uparrow}(r)=c_{1}\frac{e^{-ik_{F}r}}{\sqrt{r}}e^{-r/\xi_{0}}+c_{2}%
\frac{e^{ik_{F}r}}{\sqrt{r}}e^{-r/\xi_{0}} \label{Uup}%
\end{equation}%
\begin{equation}
u_{\downarrow}(r)=-i(c_{1}\frac{e^{-ik_{F}r}}{\sqrt{r}}-c_{2}\frac{e^{ik_{F}%
r}}{\sqrt{r}})e^{-r/\xi_{0}} \label{Udown}%
\end{equation}
where $\xi_{0}=\frac{v_{F}}{\xi_{0}}$, $k_{F}=\frac{\mu}{v_{F}}$. Note that
$c_{1}$ should equal to $c_{2}$ to ensure $u(r)$ is real, and then the radial
wave functions can be reduced as $u_{\uparrow/\downarrow}(r)\propto J_{\mp
1/2}(k_{F}r)e^{-r/\xi_{0}}$, by contrast with the vortex condition the order
of Bessel function has a $\frac{1}{2}$ shift. Note that we have obtain one
zero energy mode solution as shown in Eqs. \ref{ZMS-1}, \ref{Uup} and
\ref{Udown}. However, the Hamiltonian in Eq. \ref{1} has the time-reversal
symmetry (TRS) when $B_{z}=0$. Therefore, there must exist another zero-energy
mode, which is the TRS partner of the first one denoted by Eqs. \ref{ZMS-1},
\ref{Uup} and \ref{Udown}. Both of them form the Kramers degenerate states.
The TRS operator takes the form $\mathcal{T}=i\tau_{0}\sigma_{y}K$.

When the magnetic field $\mathbf{B}=(B_{x},B_{y},B_{z})$ is taken into
account, the Hamiltonian for the zero modes in Eq. \ref{radiu} can be
expressed as%

\begin{equation}
H_{zero}=(-i\sigma_{y}\hat{T}(r)+\alpha B_{x}\sigma_{x}+\alpha B_{y}\sigma
_{y}+\alpha B_{z}\sigma_{z})\tau_{z}+\Delta(r)\tau_{x} \label{Hzero-1}%
\end{equation}
Here, $\hat{T}(r)=$ $v_{F}(\partial_{r}+\frac{1}{2r})$. In such a case, one
can define a mirror symmetry $\mathcal{M}_{l=0}=i\sigma_{y}\tau_{y}\hat{O}%
(r)$. $\hat{O}(r)$ is a spacial inverse operator along the radial direction
with the inverse center at $R_{0}$. Note that $\hat{T}(r)$ and $\Delta(r)$
changes sign under the operation $\hat{O}(r)$. Now, one can get $[\mathcal{M}%
_{l=0},H_{zero}]=0$ for non-zero $B_{x}$ and $B_{z}$. For non-zero $B_{y}$,
$[\mathcal{M}_{l=0},H_{zero}]\neq0$. Therefore, a pair of zero modes are
robust against non-zero $B_{x}$ and $B_{z}$, but are fragile for non-zero
$B_{y}$. When $B_{z}\neq0$, we set $m_{z}=\alpha B_{z}$, the Zeeman term has
the form of $m_{z}\tau_{0}\sigma_{z}$, and the spliting between two zero
energy modes can be estimated by%

\begin{equation}
\left\langle \psi_{1}(r,\theta)|m_{z}\tau_{0}\sigma_{z}|\psi_{2}%
(r,\theta)\right\rangle =0 \label{coup-1}%
\end{equation}
Therefore, the $z$-direction Zeeman coupling cannot split the zero-energy
mode. However, if we directly solve the Eq. \ref{eeq-1} and \ref{eeq-2} with
$B_{z}\neq0$, one can find that only parameter $k_{F}$ is modified as
$k_{F}=\frac{\sqrt{\mu^{2}-m_{z}^{2}}}{v_{F}}$. It is easy to find that if
$|m_{z}|>|\mu|$ the wave function is divergent when $r\rightarrow\infty$.
Actually, only when $\mu>m_{z}$ it's topological nontrivial. \begin{figure}[h]
\centering
\par
\includegraphics[width=60mm]{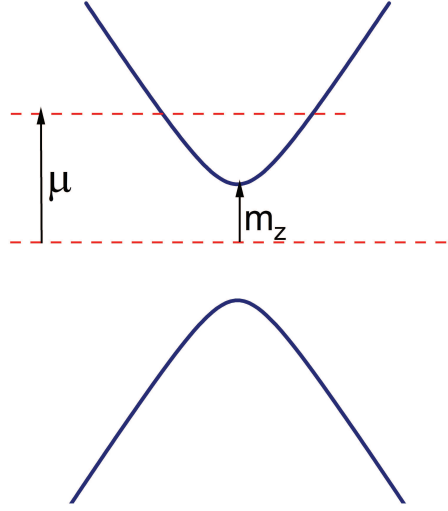}\newline\caption{TI with a z-direction
magnetic field. Only when $\mu>m_{z}$ there exists a spin-momentum-locked
fermi surface and it is topological nontrivial.}%
\label{mu}%
\end{figure}\begin{figure}[hptbh]
\centering
\par
\includegraphics[width=160mm]{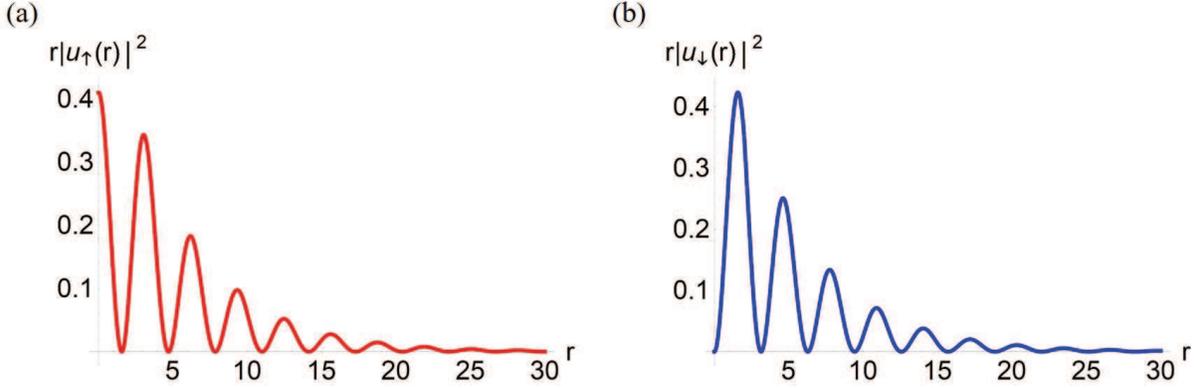}\newline\caption{Wave functions of
analytic solution.}%
\label{anay-wave}%
\end{figure}

\subsection{Quasiparticle's scattering}

\begin{figure}[h]
\centering
\par
\includegraphics[width=120mm]{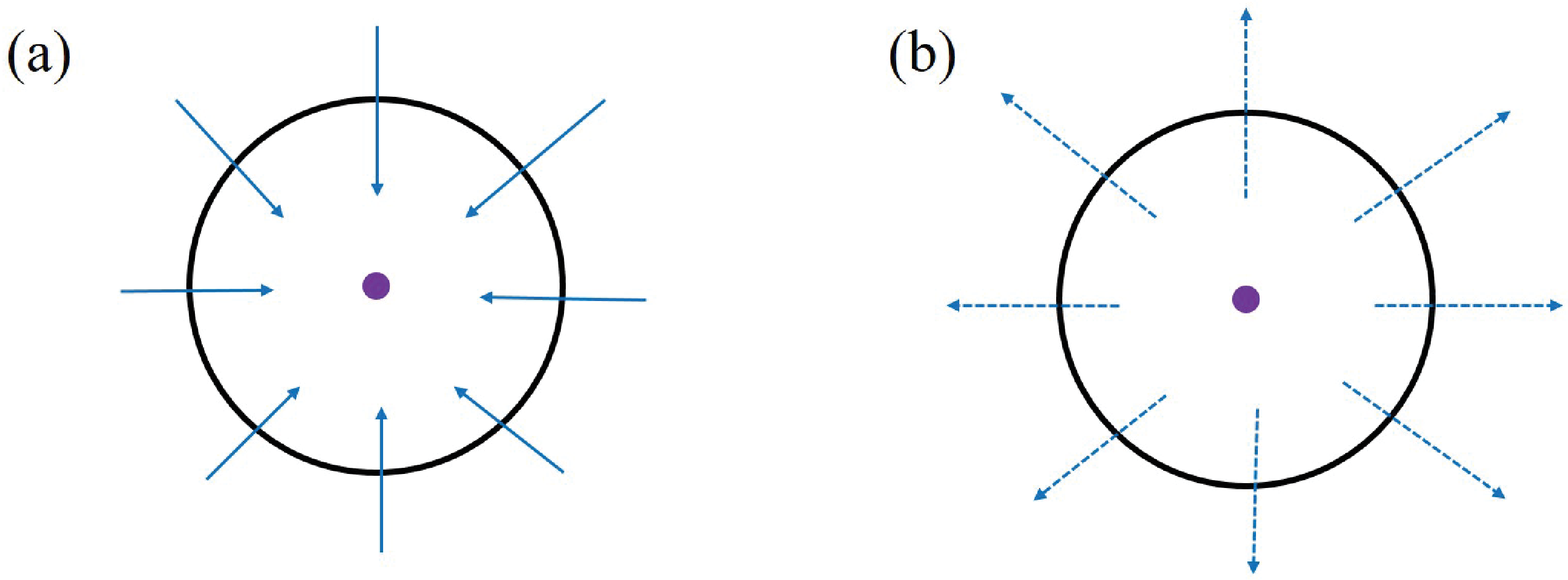} \caption{(a) The converging
cylindrical wave propagates to the impurity and (b) is scattered into a
diverging cylindrical wave. }%
\label{scattering}%
\end{figure}From Eq. (\ref{Uup}) and (\ref{Udown}), it's easy to find that the
wave functions of Majorana modes can always be divided into two parts
according to its radial propagating directions, i.e. towards or dorsad the
impurity center. We define $e^{-i k_{F}r}$ as a converging wave. Here the
scattering effect caused by the impurity should be considered. In a 2D space
where the rotation symmetry is preserved, the scattering wave function should
be described by the cylindrical wave. The scattering process can be understood
as Fig. \ref{scattering}, we name the wave functions of incoming and
scattering as $\psi_{in}$ and $\psi_{s}$, respectively, their specific form is
:
\begin{align}
\psi_{in}  &  = C \frac{e^{-i k_{F}r}}{\sqrt{r}}e^{r/\xi_{0}} \ \ \ (r<
R_{0})\\
\psi_{s}  &  = f(\theta)\frac{e^{i k r}}{\sqrt{r}},
\end{align}
where $f(\theta)$ is the scattering amplitude, and the scattering interface is
defined as $\sigma(\theta) = |f(\theta)|^{2}$ \cite{quantum}. The incoming and
scattering current density can be defined by the following formula
\begin{equation}
\label{current}j=\frac{i \hbar}{2m}(\psi\nabla\psi^{*}-\psi^{*}\nabla\psi).
\end{equation}
So the specific expression of the scattering wave function should be solved
from the flow conservation and Schr$\ddot{o}$dinger equation :
\begin{align}
j_{in}  &  = j_{s}\\
\frac{\hbar^{2}}{2m}\nabla^{2}\psi_{s}+U(r)\psi_{s}  &  = E\psi_{s}%
\end{align}
where $U(r)$ is the scattering potential, for a single impurity we can
consider it as a local potential $U(r)=U_{0}\delta(r)$, and the elastic
scattering requests $E=0$ since the scattered particle is Majorana fermion.
According to the method of partial, we can decompose the scattering wave
function into different angular-momentum channel by using
\begin{equation}
e^{ikr \cos\theta}=J_{0}(kr)+\sum\limits_{n=1}^{\infty}i^{n}J_{n}(kr)\cos
n\theta
\end{equation}
It is known that for a local $\delta$ potential, only s wave i.e. $l=0$ is
involved, so it's obvious to simplify the scattering wave function as
$\psi_{s}(r,\theta)=f(\theta)J_{0}(kr) / \sqrt{r}$. Thus the additional
density of states (DOS) at zero energy caused by the elastic scattering is
\begin{equation}
N_{s}(E=0, r, \theta)\propto r|\psi_{s}(r,\theta)|^{2} \propto J_{0}^{2}(kr).
\end{equation}
Here $|f(\theta)|$ is approximated to a constant because $\int_{0}^{2\pi
}f(\theta)d\theta=constant$. Considering this modulation about zero-energy
DOS, we obtain
\begin{equation}
\label{total-dos}N(E=0,r)=N_{s}^{0}J_{0}^{2}(kr)+N_{M}^{0}(r)
\end{equation}

\begin{figure}[h]
\centering
\par
\includegraphics[width=140mm]{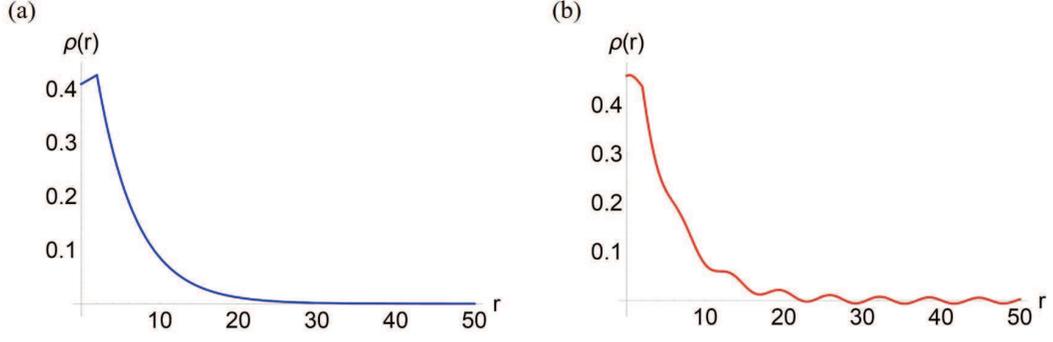} \caption{ Probability density (a)
before and (b) after taking into account of the quasiparticle-scattering
effect, the former is a simple summation of spin-up and spin-down part in Fig.
\ref{anay-wave} and the latter include the scattering wave functions.}%
\label{modulation}%
\end{figure}

\subsection{Numerical Solution}

Now we use Bessel functions as a complete orthogonal base to expand the wave
function \cite{numerical-1}. That is
\begin{align}
u_{\uparrow}(r)  &  =\sum_{n=1}^{N}u_{n}^{\uparrow}\varphi_{\nu,n}(r)\\
u_{\downarrow}(r)  &  =\sum_{n=1}^{N}u_{n}^{\downarrow}\varphi_{\nu+1,n}(r)\\
v_{\downarrow}(r)  &  =\sum_{n=1}^{N}v_{n}^{\downarrow}\varphi_{\nu+1,n}(r)\\
v_{\uparrow}(r)  &  =\sum_{n=1}^{N}v_{n}^{\uparrow}\varphi_{\nu,n}(r)
\end{align}
where $\nu=l-\frac{1}{2}$, $\varphi_{\nu,n}=\frac{\sqrt{2}}{RJ_{\nu+1}%
(j_{\nu,n}\frac{r}{R})}J_{\nu}(j_{\nu,n}\frac{r}{R})$, $N$ is the cutoff
number. Then the radial equation Eq. \ref{radiu} can be reduced as a
$4N\times4N$ matrix and the its eigenvalues are the energies. We define:
\begin{align}
T_{ij}^{\nu}  &  =\alpha\int_{0}^{R}B_{z}\varphi_{\nu,i}(r)\varphi_{\nu
,j}(r)rdr\\
V_{ij}^{\nu,\nu+1}  &  =v_{F}\int_{0}^{R}r\varphi_{\nu,i}(r)(\partial
_{r}+\frac{\nu+1}{r})\varphi_{\nu+1,j}(r)dr\\
S_{ij}^{\nu+1,\nu}  &  =v_{F}\int_{0}^{R}r\varphi_{\nu+1,i}(r)(\partial
_{r}-\frac{\nu}{r})\varphi_{\nu,j}(r)dr\\
\Delta_{ij}^{\nu}  &  =\int_{0}^{R}\Delta(r)\varphi_{\nu,i}(r)\varphi_{\nu
,j}(r)rdr
\end{align}
Here, the order parameter varies in the space can be approximately calculated
as $\Delta(r)=\Delta_{0}(1-\alpha\frac{1-\cos(2k_{F}r)}{k_{F}^{2}r^{2}})$.
Then the eigenvalue equation can be expressed as :
\begin{equation}
Det\left(
\begin{array}
[c]{cccc}%
T^{\nu}-\mu & -V^{\nu,\nu+1} & \Delta^{\nu} & 0\\
S^{\nu+1,\nu} & -T^{\nu+1}-\mu & 0 & \Delta^{\nu+1}\\
\Delta^{\nu} & 0 & T^{\nu}+\mu & V^{\nu+1,\nu}\\
0 & \Delta^{\nu+1} & -S^{\nu,\nu+1} & -T^{\nu+1}+\mu
\end{array}
\right)  =E \label{eigen-energy}%
\end{equation}
To get the numerical results, we have set the radius of the disk as
$R=800\ \mathring{A}$, the Fermi velocity $v_{F}=200\ meV\mathring{A}$ as the
experimental measurement \cite{velocity}, $\mu=1\ meV$ and the cut-off number
$N=50$ which is accurate enough to affirm the Majorana zero-energy mode. The
results are shown in Fig. \ref{num-res}.

\begin{figure}[h]
\centering
\par
\includegraphics[width=160mm]{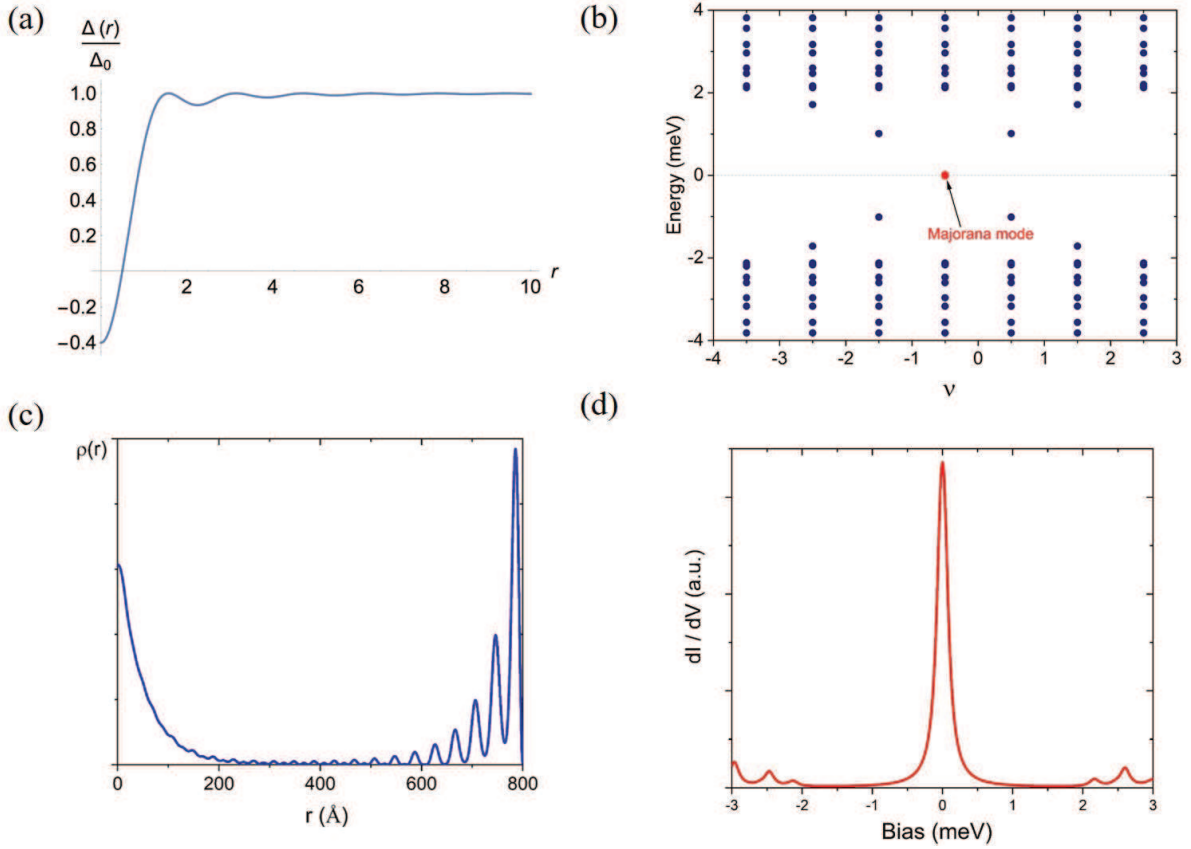}\newline\caption{(a) Variation of
order parameter in space; (b) Energy spectrum for $\nu=-\frac{1}{2} (l=0)$;
(c) Radial probability density; (d) Simulated STM spectrum at the core $r=0$.}%
\label{num-res}%
\end{figure}

The Majorana condition requests that $\mathcal{C}\psi=\eta\psi$, in the main
text we choose $\eta=-1$ because we adopt a infinite large diameter of the
disk and if $\eta=1$ the wave functions have an exponentially increasing form,
which is not physical. However if the diameter is finite, $\eta=1$ would be
reasonable and forms another probability-density peak at the boundary. Thus
the zero-energy solution should be a linear combination of $\eta=\pm1$ as
\begin{equation}
\psi=\sum_{\eta}a_{\eta}\varphi_{\eta}.
\end{equation}
And that is the reason why every zero-energy mode contains both the core and
edge state.

In the line-type $0-\pi$ junction, the gapless bound state has energy
dispersion as $E=\pm v_{F}k_{y}$. When the line is bent to form a ring, the
energy becomes discrete and can be estimated by the Bohr-Sommerfeld quantized
condition, which is :
\begin{align}
&  E_{n} = \pm v_{F}k_{n}\\
&  k_{n} 2\pi R_{0} = 2n\pi\ (n=0,1,2,\cdot\cdot\cdot)
\end{align}
Thus the mini gap is $\Delta E =v_{F}/R_{0}$. With a large Fermi velocity and
a small radius of the $\pi$-phase area, the bound states except zero-energy
Majorana mode are hidden in the ground state.

\end{widetext}


\begin{thebibliography}{99}                                                                                               %


\bibitem {YSR-1}L. Yu, Bound state in superconductors with paramagnetic
impurities, Acta Phys. Sin. \textbf{21}, 75--91 (1965).

\bibitem {YSR-2}H. Shiba, Classical spins in superconductors. Prog. Theor.
Phys. \textbf{40}, 435--451 (1968).

\bibitem {YSR-3}A. I. Rusinov and P. M. Z. E. T. Fiz, On the theory of gapless
superconductivity in alloys containing paramagnetic impurities, JETP Lett.
\textbf{9}, 1101--1106 (1968).

\bibitem {Kando-1}J. Kondo, Resistance minimum in dilute magnetic alloys,
Prog. Theor. Phys. 32, 37 (1964).

\bibitem {RMP-1}A. V. Balatsky, I. Vekhter, and Jian-Xin Zhu, Impurity-induced
states in conventional and unconventional superconductors, Re. Mod. Phys.
\textbf{78}, 373-433 (2006).

\bibitem {MZM-1}J-X. Yin, Z. Wu, J-H. Wang, Z.-Y. Ye, J. Gong, X.-Y. Hou, L.
Shan, A. Li, X.-J. Liang, X.-X. Wu, J. Li, C.-S. Ting, Z.-Q.Wang, J.-P. Hu,
P.-H. Hor, H. Ding and S. H. Pan, Observation of a robust zero-energy bound
state in iron-based superconductor Fe(Te,Se), Nat. Phys. \textbf{11}, 543 (2015)

\bibitem {MZM-2}C. Liu, C. Chen, X. Liu, Z. Wang, Y. Liu, S. Ye, Z. Wang, J.
Hu and Jian Wang, Zero-energy bound states in the high-temperature
superconductors at the two-dimensional limit, Sci. Adv. \textbf{6}, eaax7547 (2020)

\bibitem {MZM-3}P. Fan, F. Yang, G. Qian, H. Chen, Y.-Y. Zhang, G. Li, Z.
Huang, Y. Xing, L. Kong, W. Liu, K. Jiang, C. Shen, S. Du, J. Schneeloch, R.
Zhong, G. Gu, Z. Wang, H. Ding and H.-J. Gao, Observation of magnetic
adatom-induced Majorana vortex and its hybridization with field-induced
Majorana vortex in an iron-based superconductor, Nat. Commun. \textbf{12},
1348 (2021).

\bibitem {NMZM-1}D. Wang, J. Wiebe, R. Zhong, G. Gu, and R. Wiesendanger,
Spin-Polarized Yu-Shiba-Rusinov States in an Iron-Based Superconductor, Phys.
Rev. Lett. \textbf{126}, 076802 (2021).

\bibitem {NMZM-2}D. Chatzopoulos, D. Cho, K. M. Bastiaans, G. O. Steffensen,
D. Bouwmeester, A. Akbari, G. Gu, J. Paaske, B. M. Andersen and M. P. Allan,
Spatially dispersing Yu-Shiba-Rusinov states in the unconventional
superconductor FeTe$_{0.55}$Se$_{0.45}$, Nat. Commun. \textbf{12}, 298 (2021).

\bibitem {SM}See the Supplemental Material At xxx for details about the DFT
calculations, the self-consistent solutions of the BdG equations in lattice
model, the proof for the quivalance between vortex case and impurity case, and
the solutions for the majorana zero modes.

\bibitem {Neutron}V. Thampy, J. Kang, J. A. Rodriguez-Rivera, W. Bao, A. T.
Savici, J. Hu, T. J. Liu, B. Qian, D. Fobes, Z. Q. Mao, C. B. Fu, W. C. Chen,
Q. Ye, R. W. Erwin, T. R. Gentile, Z. Tesanovic, and C. Broholm, Friedel-Like
Oscillations from Interstitial Iron in Superconducting Fe$_{1+y}$Te$_{0.62}%
$Se$_{0.38}$, Phys. Rev. Lett. \textbf{108}, 107002 (2012).

\bibitem {TS-1}N. Hao and J. Hu, Topological Phases in the Single-Layer FeSe.
Phys. Rev. X \textbf{4}, 031053 (2014).

\bibitem {TS-2}Z.Wang, P. Zhang, G. Xu, L. K. Zeng, H. Miao, X. Xu, T. Qian,
H. Weng, P. Richard, A. V. Fedorov, H. Ding, X. Dai, and Z. Fang, Topological
nature of the FeSe0.5Te0.5 superconductor. Phys. Rev. B \textbf{92}, 115119 (2015).

\bibitem {TS-3}X. Wu, S. Qin, Y. Liang, H. Fan, and J. Hu, Topological
characters in Fe(Te$_{1-x}$Se$_{x}$) thin films. Phys. Rev. B \textbf{93},
115129 (2016).

\bibitem {TS-4}G. Xu, B. Lian, P. Tang, X.-L. Qi, and S.-C. Zhang, Topological
Superconductivity on the Surface of Fe-Based Superconductors. Phys. Rev. Lett.
\textbf{117}, 047001 (2016).

\bibitem {TS-5}N. Hao and J. Hu, Topological quantum states of matter in
iron-based superconductors: from concept to material realization. Natl. Sci.
Rev. \textbf{6}, 213 (2019).

\bibitem {Po-2}D. Wang, L. Kong, P. Fan, H. Chen, S. Zhu, W. Liu, L. Cao, Y.
Sun, S. Du, J. Schneeloch, R. Zhong, G. Gu, L. Fu, H. Ding, H.-J. Gao,
Science, \textbf{362}, 333-335 (2018).

\bibitem {AV-1}K. Jiang, X. Dai, and Z. Wang, Quantum Anomalous Vortex and
Majorana Zero Mode in Iron-Based Superconductor Fe(Te,Se), Phys. Rev. X
\textbf{9}, 011033 (2019).

\bibitem {Po-6}X. Wu, S. B. Chung, C. Liu, E. Kim, Phys. Rev. Research
\textbf{3}, 013066 (2021).

\bibitem {Po-7}C. Chiu and Z. Wang, Phys. Rev. Lett. \textbf{128}, 237001 (2022).

\bibitem {Po-8}Z. Zhou and J. Klinovaja, arXiv:2109.08200 (2021).

\bibitem {QPT-1}M. E. Flatt\'{e} and J. M. Byers, Local Electronic Structure
of a Single Magnetic Impurity in a Superconductor, Phys. Rev. Lett.
\textbf{78}, 3761 (1997).

\bibitem {QPT-2}R. K\H{u}mmel, Electronic Structure of Superconductors with
Dilute Magnetic Impurities, Phys. Rev. B. \textbf{6}, 2617 (1972).

\bibitem {Order-1}A. Yazdani, B. A. Jones, C. P. Lutz, M. F. Crommie, and D.
M. Eigler, Probing the Local Effects of Magnetic Impurities on
Superconductivity, Science, \textbf{275} (5307) (1997).

\bibitem {Order-2}M. I. Salkola, A. V. Balatsky, and J. R. Schrieffer,
Spectral properties of quasiparticle excitations induced by magnetic moments
in superconductors, Phys. Rev. B \textbf{55}, 12648 (1997).

\bibitem {Order-3}T. Meng, J. Klinovaja, S. Hoffman, P. Simon, and D. Loss,
Superconducting gap renormalization around two magnetic impurities: From Shiba
to Andreev bound states, Phys. Rev. B \textbf{92}, 064503 (2015).

\bibitem {MZM1D-1}L. Fu and C. L. Kane, Superconducting Proximity Effect and
Majorana Fermions at the Surface of a Topological Insulator, Phys. Rev. Lett.
\textbf{100}, 096407 (2008).

\bibitem {MZM1D-2}R. Song, P. Zhang and N. Hao, Phase-Manipulation-Induced
Majorana Mode and Braiding Realization in Iron-Based Superconductor Fe(Te,Se),
Phys. Rev. Lett. \textbf{128}, 016402 (2022).

\bibitem {Anti}Wen-Yu Shan, Jie Lu, Hai-Zhou Lu, and Shun-Qing Shen,
Vacancy-induced bound states in topological insulators, Phys. Rev. B
\textbf{84}, 035307 (2011).

\bibitem {Anti-1}Xiao-Liang Qi, Taylor L. Hughes, S. Raghu, and Shou-Cheng
Zhang, Time-Reversal-Invariant Topological Superconductors and Superfluids in
Two and Three Dimensions Phys. Rev. Lett. \textbf{102}, 187001 (2009).

\bibitem {Po-1}J. R. Colbert and P. A. Lee, Proposal to measure the
quasiparticle poisoning time of Majorana bound states, Phys. Rev. B
\textbf{89}, 140505(R) (2014).

\bibitem {Po}Fan Zhang, C. L. Kane, and E. J. Mele, Time-Reversal-Invariant
Topological Superconductivity and Majorana Kramers Pairs, Phys. Rev. Lett.
\textbf{111}, 056402 (2013).

\bibitem {PbTaSe}S. S. Zhang, J.-X. Yin, G. Dai, L. Zhao, T.-R. Chang, N.
Shumiya, K. Jiang, H. Zheng, G. Bian, D. Multer, M. Litskevich, G. Chang, I.
Belopolski, T. A. Cochran, X. Wu, D. Wu, J. Luo, G. Chen, H. Lin, F.-C. Chou,
X. Wang, C. Jin, R. Sankar, Z. Wang, and M. Z. Hasan, Phys. Rev. B
\textbf{101}, 100507(R) (2020).

\bibitem {Pa-1}P. Zhang, K. Yaji, T. Hashimoto, Y. Ota, T. Kondo, K. Okazaki,
Z. Wang, J. Wen, G. D. Gu, H. Ding, and S. Shin, Science \textbf{360}, 182 (2018).

\bibitem {MF-1}T. Choi, W. Paul, S. Rolf-Pissarczyk, A. J. Macdonald, F. D.
Natterer, K. Yang, P. Willke, C. P. Lutz, and A. J. Heinrich, Atomic-scale
sensing of the magnetic dipolar field from single atoms, Nat. Nanotechnol.
\textbf{12}, 420 (2017).

\bibitem {MF-2}T.g Choi, C. P. Lutz, A. J. Heinrich, Studies of magnetic
dipolar interaction between individual atoms using ESR-STM, Current Applied
Physics, \textbf{17}, 11, (2017).
\end{thebibliography}

\begin{thebibliography}{99}                                                                                               %


\bibitem {vasp1}G. Kresse and J. Furthm\"{u}ller, Phys. Rev. B \textbf{54},
11169 (1996).

\bibitem {vasp2}G. Kresse and D. Joubert, Phys. Rev. B \textbf{59}, 1758 (1999).

\bibitem {PAW}P. E. Blchl, Phys. Rev. B \textbf{50}, 17953 (1994).

\bibitem {PBE}J. P. Perdew, K. Burke, and M. Ernzerhof, Phys. Rev. Lett.
\textbf{77}, 3865 (1996).

\bibitem {RMP}A. I. Buzdin, Rev. Mod. Phys. \textbf{77}, 935 (2005).

\bibitem {EXP}P. Fan et al., Nat. Commun. \textbf{12}, 1348 (2021)

\bibitem {numerical-2}Jian-Xin Zhu, Wonkee Kim, C. S. Ting, and J. P.
Carbotte, Quasiparticle States around a Nonmagnetic Impurity in a
d-Density-Wave State of High-T$_{c}$ Cuprates, Phys. Rev. Lett. \textbf{87},
197001 (2001).

\bibitem {twoband}S. Raghu et al., PRB, \textbf{77}, 220503(R) (2008)

\bibitem {spin-reso-theo}Takuto Kawakami and Xiao Hu, Evolution of Density of
States and a Spin-Resolved Checkerboard-Type Pattern Associated with the
Majorana Bound State. Phys. Rev. Lett. \textbf{115}.177001 (2015).

\bibitem {spin-reso-exp}Hao-Hua Sun, Kai-Wen Zhang, et al, Majorana Zero Mode
Detected with Spin Selective Andreev Reflection in the Vortex of a Topological
Superconductor. Phys. Rev. Lett. \textbf{116}.257003 (2016).

\bibitem {quantum}Sakurai J J, Advanced Quantum Mechanics[M], Addison Wesley,
New York, (1967).

\bibitem {numerical-1}Li Mao and Chuanwei Zhang, Robustness of Majorana modes
and minigaps in a spin-orbit-coupled semiconductor-superconductor
heterostructure, Phys. Rev. B \textbf{82}, 174506 (2010).

\bibitem {velocity}D.Wang, L. Kong, P. Fan, H. Chen, S. Zhu, W. Liu, L. Cao,
Y. Sun, S. Du, J. Schneeloch, R. Zhong, G. Gu, L. Fu, H. Ding, and H.-J. Gao,
Evidence for Majorana bound states in an iron-based superconductor, Science
\textbf{362}, 333 (2018).
\end{thebibliography}
\end{document}